\documentclass{article}
\usepackage{graphicx} 
\usepackage{tabularx}
\usepackage{multicol}
\usepackage{booktabs}
\usepackage{multirow}
\usepackage{verbatim}
\usepackage[margin=1in]{geometry}
\usepackage{url}
\usepackage{listings}
\usepackage{xcolor}

\usepackage{titlesec}
\graphicspath{ {./figs/} }
\usepackage{tikz}
\usepackage[many]{tcolorbox}    	

\usepackage{booktabs}
\usepackage{xltabular}
\usepackage{array}
\usepackage{makecell}

\renewcommand{\arraystretch}{0.96}
\newcolumntype{L}[1]{>{\RaggedRight\arraybackslash}p{#1}}
\newcolumntype{Y}{>{\RaggedRight\arraybackslash}X}

\usepackage[backend=biber,style=numeric]{biblatex} 
\usepackage{tcolorbox}
\tcbuselibrary{listings, breakable}
\usepackage{listings}
\usepackage{pdfpages}
\usepackage[hidelinks]{hyperref}

\begin{document}
\begin{titlepage}
\title{Detecting Malicious Entra OAuth Apps with LLM-Based Permission Risk Scoring}
\author{Ashim Mahara}
\date{December 2025}
\thispagestyle{empty}
\end{titlepage}
\maketitle
\thispagestyle{empty}


\newpage
\begin{center}
    \thispagestyle{empty}
    \vspace*{\fill}
    To my beloved parents, sister, friends and well-wishers
    \vspace*{\fill}
\end{center}

\newpage
\section*{\centering Acknowledgements}

\begin{center}
    \thispagestyle{empty}
    \vspace*{\fill}
    I would like to extend my deepest gratitude to my advisors Prof. Nidhi Rastogi and Dipkamal Bhusal. 
    Prof. Rastogi has rooted for me even before I joined RIT. She welcomed me into her lab with open arms (and at times even fought for my presence there). Without her blessings this project, or my other endeavors, would not have been possible. Dipkamal has probably been one of my most vocal supporters in academics and real-life. His affection for me and my work have have helped me through difficult times; for which I would forever be indebted to him. 
    Prof. Wright provided me with the resources to complete this project and other projects throughout the second-half of my degree. The access that he provided me to RIT's Research Computing cluster and GPUs have enabled me to achieve these results; which would have at best times be impossible to materialize. A special thanks to everyone at Research Computing for completing my (at times absurd) requests with swiftness.
    \vspace*{\fill}
\end{center}

\newpage
\thispagestyle{empty}
\begin{center}
    \vspace*{\fill}
        \textit{Je est un autre}
    \vspace*{\fill}
\end{center}
\newpage
\thispagestyle{empty}

\tableofcontents
\thispagestyle{empty}

\newpage
\listoffigures
\thispagestyle{empty}

\newpage
\listoftables
\thispagestyle{empty}

\setcounter{page}{1}

\section{Introduction}

Open Authentication (OAuth) has emerged as the predominant protocol for authentication and authorization, enabling third-party applications to access protected resources without directly handling user credentials \cite{HowHackersAbuseOAuth_2024}. While OAuth reduces development overhead by abstracting authentication workflows and enhances user experience through centralized credential management, its widespread adoption has introduced significant attack vectors for malicious actors \cite{Bogati_2025, HowAttackersUseCompromisedAccountstoCreateandDistributeMaliciousOAuthApps|ProofpointUS_2021, HowHackersAbuseOAuth_2024, Intelligence_2022, mindgrub_2025, OAuthAttacksTargetMicrosoft365GitHub}.

The inherent trust established through user consent grants within OAuth workflows has been systematically exploited for organizational infiltration and lateral movement across enterprise networks \cite{HowAttackersUseCompromisedAccountstoCreateandDistributeMaliciousOAuthApps|ProofpointUS_2021}. Following initial compromise, threat actors frequently leverage their elevated privileges to register malicious applications with excessive permissions, enabling further exploitation. When targeting reputable organizations, adversaries weaponize the compromised entity's credibility to distribute malicious applications under the guise of legitimate software, thereby increasing the success rate of subsequent attacks.

Despite substantial documentation of OAuth application abuse in academic literature, industry reports, and technical publications \cite{Bogati_2025, HowAttackersUseCompromisedAccountstoCreateandDistributeMaliciousOAuthApps|ProofpointUS_2021, HowHackersAbuseOAuth_2024, Intelligence_2022, mindgrub_2025, OAuthAttacksTargetMicrosoft365GitHub, 6MonthsofResearchingOAuthApplicationAttacks|Huntress, WeaponizedGoogleOAuthTriggersMaliciousWebSocket_2025}, a significant gap persists in available datasets and security control resources for OAuth-based access control policies, particularly within Microsoft Azure / Entra ecosystems.

This project \footnote{https://github.com/ashim-mahara/OMGPermissions} systematically classifies 769 permissions/scopes available in Microsoft Graph API to different risk tiers using several open-source LLMs. Furthermore, to advance detection capabilities within the security community, we also open-source an automated detection framework leveraging the generated risk scores to identify malicious application registration activities while simultaneously providing reasoning traces for each of the risk scores of the permissions for explainability.

\subsection{Motivation}

While nation state actors such as Midnight Blizzard (APT 29) \cite{mircosoft_midnight_blizzard} were already documented abusing cloud applications for privilege escalation (TA0004 \footnote{https://attack.mitre.org/tactics/TA0004/}) and defense evasion (TA0005 \footnote{https://attack.mitre.org/tactics/TA0005/}) \cite{Wiz_Midnight_Blizzard} as early as January 2024, there has been a gradual increase in sophisticated OAuth-based attacks over the past year from other threat actors as well. 

In January 2025, Cyberhaven \cite{cyberhaven} reported that an unidentified adversary had released a malicious browser extension following an OAuth phishing attack. The Cyberhaven incident was part of a larger campaign that led to compromise of at-least 35 Chrome browser extensions \cite{Silberman}. In February 2025, Microsoft Threat Intelligence team \cite{Intelligence_2025} outlined that an adversary (named Storm-2372) had been targeting governments, NGOs and multiple industries in multiple regions using device code phishing as early as August 2024. A successful exploitation led to acquisition of an access token which was then abused to perform reconnaissance and data exfiltration via Microsoft Graph. The Threat Intelligence team at Volexity \cite{volexity_russian} later revealed that more Russia-based threat actors were using the same techniques used by Storm-2372.

In March 2025, Abrams (via BleepingComputer) \cite{Abrams} reported that 12000 Github repositories were targeted with a phishing campaign designed to steal Github access tokens. Later, Proofpoint \cite{proofpoint_2} revealed in July 2025 that they had started to observe an increase in Microsoft OAuth app impersonation campaigns with the success rate exceeding 50\% and attackers compromising more than 3000 user accounts across 900 Microsoft environments. Recently in October 2025, BleepingComputer \cite{bleeping_computer_co-phish} again reported a new attack vector named \textit{Co-Phish} that abused Microsoft Copilot Studio agents to send malicious OAuth consents via trusted Microsoft domains. 

In their Digital Defense Report 2025 \cite{digital_defense_2025}, Microsoft outlines cloud identity abuse as the fourth fastest growing cyber risk with attackers targeting systems by "deploying malicious apps, abusing legacy authetication, and evolving device code phishing and adversary-in-the-middle (AiTM) attacks".  In the same report, they also highlight that attackers are leveraging multi-stage attacks that utilize a number of techniques such as device code phishing and OAuth consent phishing for gaining access. 

It is evident that attackers are integrating OAuth consent phishing and rogue cloud applications to their trade craft and as such it is critical that applications and their permissions are audited on a regular basis. 

\subsection{Key Contributions}

Our core contributions are summarized as follows.

\begin{itemize}
    \item First, we provide a \textbf{public dataset} containing assigned risk scores for each of the 769 Microsoft Graph permissions/scopes from 8 different open-source LLMs. The dataset also contains reasoning for each of the permissions as assigned by the LLMs.
    \item Second, we provide \textbf{open-source code} for our detection framework which can be deployed in an enterprise environment, for Microsoft Entra, that detects and alerts on any applications with excessive scopes/permissions. 
\end{itemize}

\section{Background}

\subsection{Open Authentication (OAuth) Authorization Framework}

OAuth is formally defined by its specification authors as an authorization framework that enables third-party applications to obtain limited access to an HTTP service \cite{Hardt_2012}. Developed as a standardized protocol for secure delegated access, OAuth has been widely adopted as the industry standard for cross-application authentication and authorization workflows. The framework addresses a critical security concern in modern application ecosystems by enabling granular resource access without requiring the resource owner to disclose their credentials to third-party applications.

The fundamental innovation of OAuth lies in its token-based architecture, which replaces the traditional practice of credential sharing with a secure delegation mechanism. Through the strategic implementation of authorization codes and access tokens, OAuth facilitates a seamless user experience while maintaining robust security controls. Authorization codes serve as temporary, single-use credentials that are exchanged for access tokens, which in turn grant specified permissions to client applications within defined temporal and scope boundaries. The framework's design separates authentication from authorization, allowing specialized authorization servers to handle user authentication while resource servers focus on enforcing access policies based on token validation. 

\begin{figure}[h]
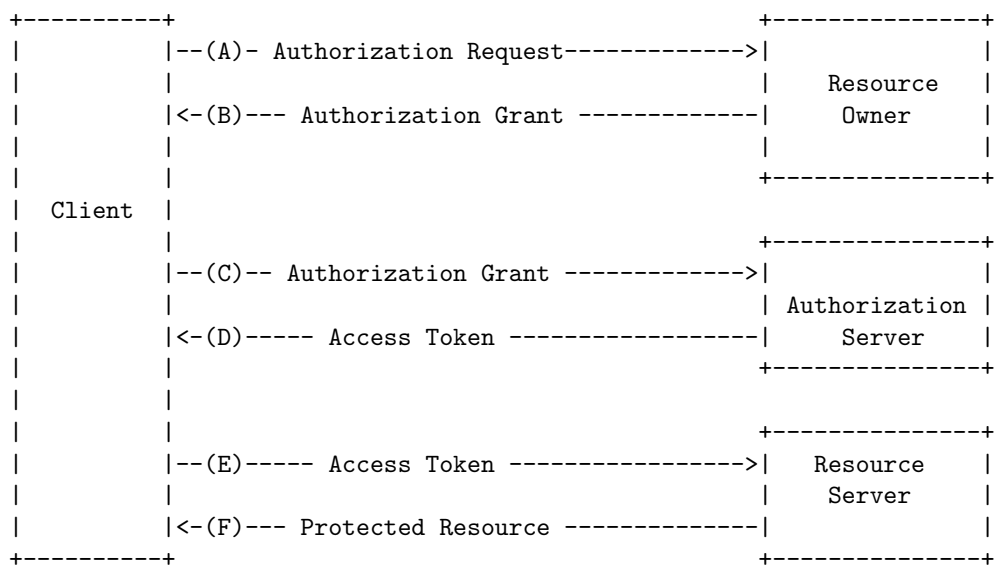

\hspace*{1cm} 
\begin{minipage}{\textwidth}
\begin{verbatim}
+----------+                                          +---------------+
|          |--(A)- Authorization Request------------->|               |
|          |                                          |    Resource   |
|          |<-(B)--- Authorization Grant -------------|     Owner     |
|          |                                          |               |
|          |                                          +---------------+
|  Client  |
|          |                                          +---------------+
|          |--(C)-- Authorization Grant ------------->|               |
|          |                                          | Authorization |
|          |<-(D)----- Access Token ------------------|     Server    |
|          |                                          +---------------+
|          |
|          |                                          +---------------+
|          |--(E)----- Access Token ----------------->|   Resource    |
|          |                                          |    Server     |
|          |<-(F)--- Protected Resource --------------|               |
+----------+                                          +---------------+
\end{verbatim}
\caption{OAuth Protocol Flow}
\label{fig:protocol-flow}
\end{minipage}
\end{figure}

\subsubsection{OAuth Workflow}

A typical OAuth workflow consists of the following entities:

\begin{itemize}
\item \textbf{Resource Owner}: The entity capable of granting access to a protected resource. Typically, this is the end-user.
\item \textbf{Resource Server}: The server hosting the protected resources, capable of accepting and responding to protected resource requests using access tokens.
\item \textbf{Client}: An application making protected resource requests on behalf of the resource owner.
\item \textbf{Authorization Server}: The server that authenticates the resource owner and issues access tokens to the client after obtaining authorization.
\end{itemize}

\subsubsection{Authorization Codes and Access Tokens}

The core OAuth 2.0 framework defines several grant types, with the Authorization Code grant being the most common for server-side web applications. This flow introduces two fundamental concepts: the Authorization Code and the Access Token.

The \textbf{Authorization Code} is a temporary credential issued to the client by the authorization server via the resource owner's user-agent. It is obtained after the resource owner has successfully authenticated and authorized the client. The authorization code is a short-lived, single-use token whose primary purpose is to be exchanged for an access token. This intermediary step enhances security by ensuring that the access token is transmitted directly from the authorization server to the client, without being exposed to the resource owner's user-agent (e.g., a web browser).

An \textbf{Access Token} is a credential used by the client to access the protected resources hosted by the resource server. It is a string representing an authorization issued to the client and is typically opaque to the client. Tokens can specify their scope and duration of access. Unlike the authorization code, the access token is presented by the client directly to the resource server when making API requests. The resource server must validate the token and ensure it has not expired and is authorized for the requested operation. \\



\subsection{OAuth 2.0 Implementation in Microsoft Entra ID}

Microsoft Entra ID (formerly Azure Active Directory) constitutes a comprehensive implementation of the OAuth 2.0 authorization framework, serving as the foundational identity and access management service within Microsoft's cloud ecosystem [\cite{OwenRichards1}, \cite{Bhaskardeep_Khaund_2025}]. While maintaining strict adherence to the OAuth 2.0 specification \cite{Hardt_2012}, Entra ID incorporates enterprise-grade enhancements that address organizational security and governance requirements. This implementation preserves the core OAuth architectural principles while optimizing them for enterprise deployment scenarios.

The conceptual mapping between standard OAuth entities and their Entra ID counterparts demonstrates the framework's adaptation to enterprise contexts:

\begin{itemize}
    \item \textbf{Resource Owner}: Organizational identities managed within Entra ID tenants, typically authenticated through enterprise credentials.
    \item \textbf{Resource Server}: Microsoft cloud services including Microsoft Graph API, SharePoint Online, Exchange Online, and other protected resources within the Microsoft cloud infrastructure.
    \item \textbf{Client}: Third-party SaaS applications or custom-developed line-of-business applications requiring access to organizational resources.
    \item \textbf{Authorization Server}: The Entra ID service itself, providing integrated authentication and authorization token management capabilities.
\end{itemize}

The OAuth authorization flow for any applications utilizing the Microsoft Identity Platform is identical (as illustrated in Fig \ref{fig:entra_oauth_flow}). 

\usetikzlibrary{arrows.meta, positioning, shapes, backgrounds}

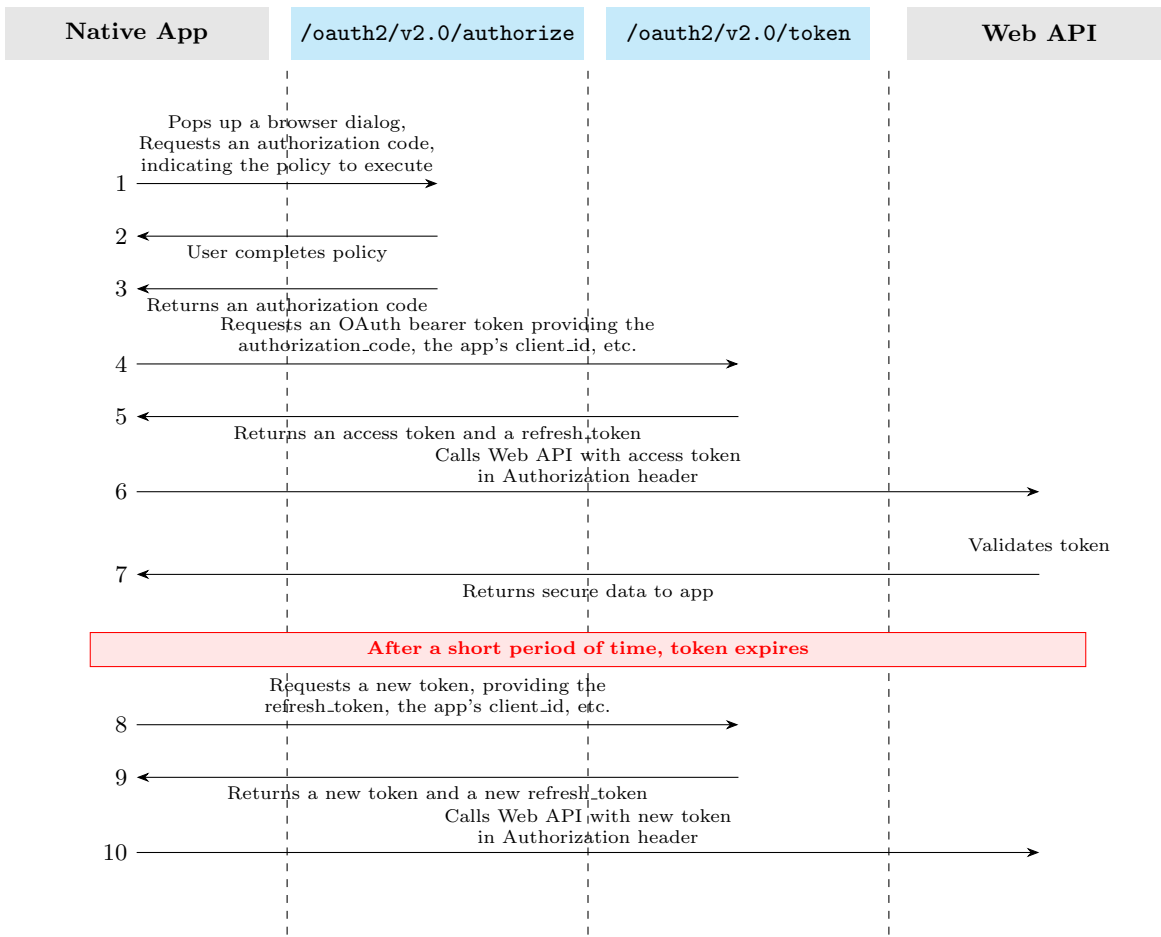
\begin{figure}[h]
\centering
\begin{tikzpicture}[>=Stealth, font=\small, node distance=0.5cm]

\def\nativeX{0}
\def\authorizeX{4}
\def\tokenX{8}
\def\apiX{12}

\node[fill=blue!20, text width=15cm, minimum height=1cm, align=center, font=\bfseries] at (6,0.5) 
    {Microsoft Identity Platform\\
    \scriptsize\url{https://login.microsoftonline.com/<tenant>} or \url{https://login.microsoftonline.com/common/}};

\node[fill=gray!20, minimum width=3.5cm, minimum height=0.7cm, align=center] at (\nativeX,-1) {\textbf{Native App}};
\node[fill=cyan!20, minimum width=3.5cm, minimum height=0.7cm, align=center] at (\authorizeX,-1) {\ttfamily /oauth2/v2.0/authorize};
\node[fill=cyan!20, minimum width=3.5cm, minimum height=0.7cm, align=center] at (\tokenX,-1) {\ttfamily /oauth2/v2.0/token};
\node[fill=gray!20, minimum width=3.5cm, minimum height=0.7cm, align=center] at (\apiX,-1) {\textbf{Web API}};

\draw[dashed] (2,-1.5) -- (2,-13);
\draw[dashed] (6,-1.5) -- (6,-13);
\draw[dashed] (10,-1.5) -- (10,-13);

\def\startY{-3}

\draw[->] (\nativeX,\startY) -- (\authorizeX,\startY) node[midway, above, align=center, font=\scriptsize] 
    {Pops up a browser dialog,\\Requests an authorization code,\\indicating the policy to execute};
\node at (\nativeX,\startY) [left] {1};

\draw[->] (\authorizeX,\startY-0.7) -- (\nativeX,\startY-0.7) node[midway, below, align=center, font=\scriptsize] 
    {User completes policy};
\node at (\nativeX,\startY-0.7) [left] {2};

\draw[->] (\authorizeX,\startY-1.4) -- (\nativeX,\startY-1.4) node[midway, below, align=center, font=\scriptsize] 
    {Returns an authorization code};
\node at (\nativeX,\startY-1.4) [left] {3};

\draw[->] (\nativeX,\startY-2.4) -- (\tokenX,\startY-2.4) node[midway, above, align=center, font=\scriptsize] 
    {Requests an OAuth bearer token providing the\\authorization\_code, the app's client\_id, etc.};
\node at (\nativeX,\startY-2.4) [left] {4};

\draw[->] (\tokenX,\startY-3.1) -- (\nativeX,\startY-3.1) node[midway, below, align=center, font=\scriptsize] 
    {Returns an access token and a refresh\_token};
\node at (\nativeX,\startY-3.1) [left] {5};

\draw[->] (\nativeX,\startY-4.1) -- (\apiX,\startY-4.1) node[midway, above, align=center, font=\scriptsize] 
    {Calls Web API with access token\\in Authorization header};
\node at (\nativeX,\startY-4.1) [left] {6};

\node[align=center, font=\scriptsize] at (\apiX,\startY-4.8) {Validates token};
\draw[->] (\apiX,\startY-5.2) -- (\nativeX,\startY-5.2) node[midway, below, align=center, font=\scriptsize] 
    {Returns secure data to app};
\node at (\nativeX,\startY-5.2) [left] {7};

\node[fill=red!10, draw=red, text=red, text width=13cm, align=center, font=\scriptsize] 
    at (6,\startY-6.2) {\textbf{After a short period of time, token expires}};

\draw[->] (\nativeX,\startY-7.2) -- (\tokenX,\startY-7.2) node[midway, above, align=center, font=\scriptsize] 
    {Requests a new token, providing the\\refresh\_token, the app's client\_id, etc.};
\node at (\nativeX,\startY-7.2) [left] {8};

\draw[->] (\tokenX,\startY-7.9) -- (\nativeX,\startY-7.9) node[midway, below, align=center, font=\scriptsize] 
    {Returns a new token and a new refresh\_token};
\node at (\nativeX,\startY-7.9) [left] {9};

\draw[->] (\nativeX,\startY-8.9) -- (\apiX,\startY-8.9) node[midway, above, align=center, font=\scriptsize] 
    {Calls Web API with new token\\in Authorization header};
\node at (\nativeX,\startY-8.9) [left] {10};

\end{tikzpicture}
\caption{OAuth2 Authorization Code Flow with Refresh Token (Microsoft Identity Platform \cite{OwenRichards1}}
\label{fig:entra_oauth_flow}
\end{figure}

\subsubsection{Resource Permissions in Entra ID}

The Entra ID implementation also extends the core OAuth specification through several enterprise-specific capabilities.

\begin{itemize}
    \item \textbf{Administrative consent workflows} \footnote{https://learn.microsoft.com/en-us/entra/identity/enterprise-apps/admin-consent-workflow-overview} enable tenant administrators to grant organization-wide application permissions, circumventing individual consent requirements for lower-risk access scenarios.
    
    \item \textbf{Application-level permissions} \footnote{https://learn.microsoft.com/en-us/entra/identity-platform/permissions-consent-overview} support server-to-server authentication scenarios without user context, subject to rigorous administrative approval processes.

    \item \textbf{Delegated permissions} \footnote{https://learn.microsoft.com/en-us/entra/identity-platform/delegated-access-primer} allow an application to act on behalf of a signed-in user to access resources that the user has access to. The application's effective permissions are the intersection of the permissions it has been granted and the privileges of the user who is signed in.
    
    \item \textbf{Permission classification framework} \footnote{https://learn.microsoft.com/en-us/entra/identity/enterprise-apps/configure-permission-classifications?pivots=portal} allows administrative categorization of API permissions according to risk impact levels, directly influencing user consent experiences and security policy enforcement. While Microsoft does not provide default permission classifications, the Security Copilot agent can be enabled (requires Security Capacity Units; additional paid Azure resources) to generate policy recommendations. 
\end{itemize}

Additionally, Entra ID provides extensive \textbf{audit and monitoring capabilities} that log all consent operations and token issuance events, enabling security operations teams to detect anomalous access patterns and potential security incidents. 

\subsection{LLMs in Cybersecurity}

Zhang et al \cite{Zhang_Bu_Wen_Liu_Fei_Xi_Li_Yang_Zhu_Meng_2025} conducted a survey of over 300 papers, published 2023 onward, focused on applying LLMs to cybersecurity. They categorized the applications to 9 specific categories including \textit{Threat Intelligence}, \textit{Vulnerability Detection}, and \textit{Malware Detection}. They conclude their research question \#2 : \textit{What are the potential applications of LLMs in cybersecurity?} with "\textit{LLM security copilot can effectively empower the automation and intelligence of cybersecurity, helping to address security risk challenges.}" 

The most recent example of adoption of LLMs in defensive cybersecurity was perhaps the 2025 AIxCC challenge by DARPA \footnote{https://aicyberchallenge.com}. The two year challenge was focused on using LLMs to discover vulnerabilities in software and generate automatic patches for them. The winning team (Team Atlanta \cite{Team_Atlantis}) won \$4 million as a cash prize and found numerous zero-day (some were unknown even to the organizers) vulnerabilities in the target software repositories. Our previous work \cite{Bhusal_Alam_Nguyen_Mahara_Lightcap_Frazier_Fieblinger_Torales_Blakely_Rastogi_2024} SECURE also explored the applications of LLMs assessing their effectiveness for LLMs in realistic cybersecurity scenarios. The findings demonstrated that leading frontier models, at the time of the paper release, were either on-par or better than humans. The introduction of Microsoft Co-pilot for Security Guided Response \cite{copilot} as an integration into their core Defender product also exemplifies how large-scale enterprise vendors are also increasingly relying on LLMs to perform incident response and improve detections across their data sources. 

LLMs also exhibit substantial effectiveness in offensive security. Mayoral-Vilches et. al \cite{Mayoral-Vilches_Navarrete-Lozano_Balassone_Sanz-Gómez_Chavez_Torres_Turiel_2025} demonstrated that their Cybersecurity AI (CAI) system repeatedly beat human teams at multiple cybersecurity events. Their autonomous agent framework achieved success in five international CTF (Capture The Flag) competitions and even won \$50000 for the Neurogrid CTF as a prize. Their framework demonstrated high solve rate at ~91\% for two of the CTFs and solved the challenges 37\% faster than human teams. The results from their active participation in the competitions - against humans competing in the same capacity and time constraints - confirm two facts: (1) significant capability gains for modern LLMs when compared to the previous generation of LLMs and their capabilities during the initial adoption phase (circa 2023) - as documented in early works such as PentestGPT \cite{pentest_gpt}, (2) LLMs are significantly faster than humans at consistently solving cybersecurity challenges that require deep expertise.

As it is evident that LLMs possess verifiable cybersecurity capabilities, in both defensive and offensive security, it is natural that their usage should be expanded into more avenues to speed up processes that require significant human time and effort - with a human operator (expert) acting as the verifier to reduce stochastic errors and ensure alignment to the task. 

\subsection{Related Work}

Farooqi et al. \cite{Farooqi_Zaffar_Leontiadis_Shafiq_2017} (circa 2017) uncovered a large-scale operation that collected OAuth access tokens from colluding members to provide fake interactions in the Facebook \footnote{https://www.facebook.com/} platform. They carried out a comprehensive study of the collusion network to identify the abuse of the Facebook Graph API to provide fake interaction services to the members of the collusion network. They proposed solutions to the token abuse as: (1) Access Token Rate Limits (2) Honeypot based Access Token Invalidation; and traditional methods such as: (3) Temporal Clustering (to identify members of the collusion network) and (4) IP and ASN based limits. While their work is vastly different from our work, they demonstrated early abuse of the OAuth mechanism.

Longo's Master's Thesis work \cite{neural_permissions} "Applying Neural Topic Modeling to Detect Anomalous  Permission Requests in Microsoft 365 Applications" is perhaps the most similar to our work. In short, they framed the problem as an anomaly detection task - map permissions requested by an app to different topics and subsequently run unsupervised anomaly detection algorithms within the identified topics. They collected permissions from 1069 Entra specific applications to build their dataset but also acquiesced that there were no ground truth labels for proper evaluation/validation and instead opted to assign risk scores to individual permissions and building labels based on the aggregated risk scores; grounded by the permission lists provided by Tenable. \footnote{https://www.tenable.com/indicators/ioe/entra/DANGEROUS-APPLICATION-PERMISSIONS-AFFECTING-DATA} \footnote{https://www.tenable.com/indicators/ioe/entra/DANGEROUS-DELEGATED-PERMISSIONS-AFFECTING-DATA} We view their work as a parallel effort that explored the possibility of using anomaly detection for the task. While they have not released their dataset yet (for comparison), we affirm that their work can be combined with the method we outline in our work to build a more robust detection framework.
\section{Threat Models and Advanced Attacks}

The robust authorization model employed by OAuth 2.0 and its Entra ID implementation introduces new attack surfaces, shifting the focus from credential theft to authorization abuse. The most prevalent of these modern threat vectors, particularly in the context of enterprise cloud environments, are consent phishing and malicious application registration \cite{MicrosoftEntraIDPentesting}. These attacks exploit the user's trust in the authorization server's legitimate interface and the administrative delegation capabilities inherent in Entra ID's application model.

\subsection{Consent Phishing}

Consent phishing represents a sophisticated social engineering attack that weaponizes the legitimate OAuth consent flow to gain delegated access to a resource owner's data. This attack differs fundamentally from traditional credential phishing, as the user is directed to the genuine Microsoft Entra ID authentication and authorization endpoint, thereby bypassing URL-based security filters \cite{elastic_research}. The core mechanism involves deceiving the user into granting excessive permissions to an attacker-controlled application.

\subsubsection{Threat Vector within Entra ID}

\begin{enumerate}
\item \textbf{Malicious Application Setup}: The adversary registers a new multi-tenant application within Entra ID with a name, logo, and publisher display that mimics a legitimate, well-known service (e.g., "OneDrive Sync," "Adobe PDF Converter," or "Zoom Integration").

\item \textbf{Phishing Campaign}: The attacker launches a phishing campaign (e.g., via email or instant messaging) containing a carefully crafted link that initiates the standard OAuth 2.0 Authorization Code flow. The link directs the user to the \texttt{https://login.microsoftonline.com} authorization endpoint with the malicious application's \texttt{client\_id} and a request for high-impact delegated permissions, such as \texttt{Mail.ReadWrite}, \texttt{Files.ReadWrite.All}, or \texttt{User.Read.All}.

\item \textbf{Authentication and Deception}: The user is presented with the legitimate Microsoft sign-in page and successfully authenticates. They are then shown the Entra ID consent screen, which displays the deceptive application name and the requested permissions. Due to the high-trust environment of the Microsoft domain and the subtle naming deception, the user is socially engineered into clicking "Accept."

\item \textbf{Access Acquisition and Persistence}: Upon acceptance, the malicious application receives a valid authorization code, which it exchanges for an access token and, crucially, a refresh token. The refresh token provides persistent, non-interactive access to the user's data and resources on the Microsoft Graph API, enabling prolonged data exfiltration or subsequent activities without the need for the user's credentials or further consent.

\end{enumerate}

The prevalence of this threat in Entra ID is amplified by the sheer scope of permissions available through the Microsoft Graph API (769 different permissions), offering extensive access to organizational assets, including email, files, calendar data, and directory information \cite{FaithOmbongi}. The successful execution of a consent phishing campaign effectively creates a persistent backdoor into the organization, relying on authorization abuse rather than authentication compromise.

\newpage

\subsection{Malicious Application Registration}

Malicious application registration, often referred to as "rogue application" creation, focuses on leveraging default tenant configurations that permit standard users to register new applications and service principals within the Entra ID environment. This capability, designed for developer agility, poses a significant risk by allowing an attacker to establish a persistent presence within the tenant following a low-privilege account compromise \cite{Woodruff_2024}.

\subsubsection{Threat Vector within Entra ID}

\begin{enumerate}
\item \textbf{Initial Compromise}: An adversary compromises a non-administrative user account (e.g., via credential stuffing or traditional phishing).
\item \textbf{Application Registration}: The attacker utilizes the compromised user's identity to call the Microsoft Graph API or the Entra ID portal to register a new application. This action creates a client application object and an associated service principal within the tenant.
\item \textbf{Credentialing and Persistence}: The attacker generates credentials (e.g., a client secret or an X.509 certificate) for the newly created application. These credentials allow the attacker to authenticate the service principal directly to Entra ID, establishing a non-user-dependent persistence mechanism. This service principal is now a permanent security identity separate from the initially compromised user account.
\item \textbf{Privilege Escalation (Optional)}: If the attacker seeks high-level permissions (\textit{Application Permissions} that affect all users, like \texttt{User.ReadWrite.All}), they may pivot to an "Admin Consent Phishing" approach or rely on a separate administrative account compromise to grant the broad organizational consent required for these scopes.
\item \textbf{Data Exfiltration}: Once the service principal is granted the necessary application permissions, the attacker can use the application's credentials to execute server-to-server API calls, accessing and exfiltrating data organization-wide (e.g., reading all mailboxes or modifying directory objects) without ongoing user authentication.
\end{enumerate}

This threat model is particularly critical because the registered application acts as a permanent backdoor; and disabling the initially compromised user account does not revoke the application's access. 

\section{Proposed Solution: Proof-of-Concept for Malicious Application Detection (with Floors, Caps, Synergy \& Spike Logic)} \label{proposed-solution}

Our method focuses on the \textbf{Initial Persistence Phase} of an OAuth attack, aiming to detect the malicious application registration and the subsequent illicit consent grant. We couple Audit Log triggers with a \emph{consent-inventory–driven} risk aggregator that operates on the actual granted permissions (scopes/roles) observed for the client application.

\subsection{Aggregated Application Risk ($\mathbf{R_{\text{app}}}$)}

Let $\mathcal{P}$ be the set of granted permissions for an application. Each permission $p\!\in\!\mathcal{P}$ has a risk score $r(p)\!\in\!\{1,\dots,5\}$ generated by an LLM. We harden aggregation with:
\begin{itemize}
    \item \textbf{Floors} $f(p)$ (regex-based): e.g., \texttt{*.ReadWrite.All}, \texttt{RoleManagement.ReadWrite.*}, \texttt{Application.ReadWrite.All}, \texttt{ServicePrincipal.ReadWrite.All}, \texttt{AppRoleAssignment.ReadWrite.All}, \texttt{SecurityActions.ReadWrite.All}, and \texttt{*Policy*ReadWrite*} $\Rightarrow f(p)=5$; \texttt{*.Read.All} and sensitive \texttt{*.Send} $\Rightarrow f(p)=4$.
    \item \textbf{Caps} $c(p)$ for constrained scopes (e.g., \texttt{Files.ReadWrite.AppFolder}, \texttt{createdByApp}) where $c(p)=2$.
\end{itemize}
Define the floor-adjusted score $s(p)=\max\{r(p),\,f(p)\}$ and the set $S=\{s(p)\mid p\in\mathcal{P}\}$. Let $M_p(S)$ be the generalized (power) mean of order $p$ (default $p=3$):
\[
M_p(S)=\Big(\frac{1}{|\mathcal{P}|}\sum_{x\in S} x^{\,p}\Big)^{\!1/p},\quad
M_0(S)=\exp\!\Big(\frac{1}{|\mathcal{P}|}\sum_{x\in S}\ln x\Big).
\]
We form a conservative base score
\[
b=\max\big(\mathrm{median}(S),\;\max(S),\;M_{p}(S),\;\max_{p\in\mathcal{P}} f(p)\big).
\]
Map $b$ to a risk tier $T$ via fixed thresholds: \texttt{critical} if $b\!\ge\!4.5$, \texttt{high} if $b\!\ge\!3.5$, \texttt{medium} if $b\!\ge\!2.0$, else \texttt{low}. Then apply:
\begin{itemize}
    \item \textbf{Synergy (persistence bump)}: if \texttt{offline\_access}$\in\mathcal{P}$ and $\exists\,x\in S:\,x\ge 3$, then bump $T$ one tier (low$\to$medium$\to$high$\to$critical).
    \item \textbf{Caps (tempering)}: if any cap $c(p)=2$ applies, no high floor was triggered ($\max f(p)<4$), and $T\in\{\texttt{high},\texttt{critical}\}$, reduce $T$ toward \texttt{low}.
\end{itemize}
For numeric reporting we use the representative
\[
R_{\text{app}}=\max\big(b,\;m(T)\big),\quad 
m(T)\in\{5.0,\,4.0,\,2.5,\,1.5\}\ \text{for}\ 
T\in\{\texttt{critical},\texttt{high},\texttt{medium},\texttt{low}\}.
\]
\subsection{Stateful Spike Logic}

To catch abrupt risk escalations, we track spikes between consecutive snapshots:
\[
S_{\ge\theta}=\{p\in\mathcal{P}\mid s(p)\ge \theta\},\quad
\texttt{count\_spike}=|S_{\ge\theta}|,\quad
\texttt{spike\_ratio}=\frac{|S_{\ge\theta}|}{|\mathcal{P}|},
\]
with default spike threshold $\theta=5$. We emit spike alerts when either:
\begin{enumerate}
    \item \textbf{First spike appears}: previously no spikes, now $|S_{\ge\theta}|\ge 1$; or
    \item \textbf{Multiple/ratio spike}: $\texttt{count\_spike}\ge 2$ \emph{or} $\texttt{spike\_ratio}\ge 0.25$, and the spike set grew.
\end{enumerate}
Spike alerts observe a per-app cool-down window to avoid alert storms. They can be configured to \emph{bypass tier thresholds} (useful for p5-critical events).


\newtcolorbox{databox}{
    fontupper=\small\ttfamily,
    boxrule=0.5pt,
    colframe=gray!50,
    colback=gray!5,
    left=8pt,
    right=8pt,
    top=6pt,
    bottom=6pt,
    arc=3pt
}

\section{LLM-Assisted Permission Risk Generation} \label{sec: llm-perm-risk-gen}

To operationalize the permission–risk mapping required by the detection pipeline, we constructed a comprehensive dataset of Microsoft Graph API permissions and evaluated each through a controlled large language model (LLM) inference process. This procedure produced a structured and explainable risk score for every unique permission entry in Microsoft Entra ID.

\subsection{Permission Corpus Construction}

All available Microsoft Graph permissions were programmatically scraped from the official documentation\footnote{https://github.com/microsoftgraph/microsoft-graph-docs-contrib/blob/main/concepts/permissions-reference.md}. We extracted \textbf{769} unique permissions from the Azure documentation for this purpose. Both \texttt{delegated} and \texttt{application} variants were retained, along with their identifiers (\texttt{appRoleId}, \texttt{oauth2PermissionId}), display names, and descriptions. 

This resulted in a raw corpus $\mathcal{C}$ of approximately $N$ unique permission definitions:
\[
\begin{aligned}
p_i = \{&\texttt{permission}, \texttt{display\_application}, \texttt{display\_delegated}, \\
       &\texttt{description\_application}, \texttt{description\_delegated}, \\
       &\texttt{admin\_consent\_application}, \texttt{admin\_consent\_delegated}, \\
       &\texttt{application\_guid}, \texttt{delegated\_guid}\}
\end{aligned}
\]

The \textit{application\_guid} and \textit{delegated\_guid} are specially useful when deciding whether the application has tenant-wide access or user-specific access. The \textit{admin\_consent\_*} fields determine whether the permission require admin consent for application level and user delegation level access.

\subsubsection{Example Permission Record}

\begin{databox}
\begin{tabbing}
permission: \= IdentityRiskyServicePrincipal.ReadWrite.All \\
application\_guid: \= cb8d6980-6bcb-4507-afec-ed6de3a2d798 \\
delegated\_guid: \= bb6f654c-d7fd-4ae3-85c3-fc380934f515 \\
display\_application: \= Read and write all identity risky service principal information \\
display\_delegated: \= Read and write all identity risky service principal information \\
description\_application: \= \parbox[t]{0.7\linewidth}{Allows the app to read and update identity risky service principal for your organization, without a signed-in user.} \\
description\_delegated: \= \parbox[t]{0.7\linewidth}{Allows the app to read and update identity risky service principal information for all service principals in your organization, on behalf of the signed-in user. Update operations include dismissing risky service principals.} \\
admin\_consent\_application: \= true \\
admin\_consent\_delegated: \= true
\end{tabbing}
\end{databox}

\subsection{Model Inference Environment}

We deployed multiple large language models using \textbf{vLLM} \cite{vllm}, an optimized high-throughput inference engine supporting batched parallel serving across GPUs. Each model was hosted locally on dedicated inference nodes (2x Nvidia GH200) provided by \textbf{Rochester Institute of Technology's Research Computing} \cite{RIT_ResearchComputing_2025} to maintain reproducibility and cost control.  
The vLLM deployment allowed concurrent evaluation of hundreds of permissions per minute, while preserving deterministic prompt structure, batched inference and caching efficiency (prefix cache reuse for the system and framework sections of the prompt).

\subsection{Prompt Design and Evaluation Criteria}

Each permission record $p_i$ was supplied to the model through a structured risk-assessment prompt designed for cybersecurity reasoning. We used the LiteLLM Python SDK \footnote{https://github.com/BerriAI/litellm} to send requests to the vLLM API endpoint. 
The prompt enforced a fixed JSON response format and required both a numeric risk score and concise justification.  
The exact template was as follows:

\newtcolorbox{boxA}{
    fontupper=\small\ttfamily,
    boxrule=0.5pt,
    colframe=gray!50,
    colback=gray!5,
    left=8pt,
    right=8pt,
    top=6pt,
    bottom=6pt,
    arc=3pt
}

\begin{boxA}\label{prompt-v0}
\small
You are a cybersecurity risk assessment specialist. Analyze the provided Microsoft Graph API permission and assign a comprehensive risk score from 1-5, where:

\begin{itemize}
    \item \textbf{1:} Minimal risk (public data, read-only)
    \item \textbf{2:} Low risk (limited read access)
    \item \textbf{3:} Moderate risk (broad read or limited write access)
    \item \textbf{4:} High risk (sensitive data write access)
    \item \textbf{5:} Critical risk (full administrative control)
\end{itemize}

\subsection*{Assessment Framework}
Evaluate the permission based on these risk factors:

\begin{enumerate}
    \item \textbf{Data Sensitivity \& Scope}
    \begin{itemize}
        \item What types of data can be accessed (PII, credentials, organizational data)?
        \item Is access limited to user's own data or broader organizational access?
    \end{itemize}
    
    \item \textbf{Privilege Escalation Potential}
    \begin{itemize}
        \item Could this permission be used to gain additional privileges?
        \item Does it enable modifying security settings or user permissions?
    \end{itemize}
    
    \item \textbf{Impact of Misuse}
    \begin{itemize}
        \item What damage could occur if maliciously used?
        \item Potential for data breach, service disruption, or financial impact?
    \end{itemize}
    
    \item \textbf{Attack Vectors}
    \begin{itemize}
        \item How might attackers exploit this permission?
        \item Common misuse patterns in real-world attacks?
    \end{itemize}
\end{enumerate}

\subsection*{Response Format}
Provide a JSON object with the following structure:

\begin{verbatim}
{
    "risk_score": 5,
    "reasoning": "Concise explanation covering key risk factors"
}
\end{verbatim}

\textbf{Permission Data:}
\end{boxA}

\subsection{Automated Risk Dataset Generation}

For each permission $p_i \in \mathcal{C}$, the inference process executed the following steps:

\begin{enumerate}
    \item Constructed the JSON payload using the scraped metadata fields.
    \item Sent the structured prompt to the vLLM-served model instance.
    \item Parsed the model’s output to extract the \texttt{risk\_score} and textual \texttt{reasoning}.
    \item Stored results into a consolidated SQLite dataset (\texttt{permission\_analysis.db}), preserving model name, timestamp, and raw output for audit. \footnote{The database schema for the permission\_analysis can be found in the Appendix: Table \ref{table:db_schema_permission_analysis}}
\end{enumerate}

\subsubsection{Typical Model Response}

\begin{databox}
\begin{verbatim}
{
  "risk_score": 3,
  "reasoning": "Provides broad read access to identity providers; misuse could
  expose configuration but not alter authentication flow."
}
\end{verbatim}
\end{databox}

\subsection{Post-processing and Validation}

All generated responses were schema-validated and normalized.  
Non-numeric or incomplete outputs were discarded or re-queried.  
For stability analysis, multiple LLM variants (e.g., \texttt{Llama-3.1-8B-Instruct}, \texttt{gpt-oss}, \texttt{Qwen3}) were used on overlapping subsets to measure cross-model consistency and detect outliers.

The final curated dataset provided:
\begin{itemize}
    \item A numerical risk distribution (1–5) per permission.
    \item Corresponding rationale strings suitable for analyst review.
    \item Consistent JSON schema integration with the detection pipeline’s in-memory cache (loaded from \texttt{permission\_analysis.db}).
\end{itemize}

\subsection{LLM Models}

We generated risk scores using 8 leading open-source LLM models; interestingly, the more recent models are of Mixture-of-Experts [\cite{moe-hinton}, \cite{shazeer2017outrageouslylargeneuralnetworks}] architecture as can be seen in Table \ref{tab:model_params_moe}. \footnote{The models were selected based solely on the authors' arbitrary decision.}

\begin{table}[ht]
\centering
\small
\begin{tabular}{llll}
\toprule
Model & Params (total / active) & MoE? & Release date \\
\midrule
\texttt{granite-4.0-h-tiny} \cite{granite2025} & 7B / 1B active & Yes (hybrid MoE) & Oct 2, 2025 \\
\texttt{Llama-3.1-8B-Instruct} \cite{grattafiori2024llama3herdmodels} & ~8B & No (dense) & July 2024 (instruct release) \\
\texttt{DeepSeek-R1-Distill-Qwen-32B} \cite{deepseekai2025deepseekr1incentivizingreasoningcapability} & ~32B (dense distilled) & No (dense) & Jan 20, 2025 \\
\texttt{gpt-oss-120b} \cite{openai2025gptoss120bgptoss20bmodel} & ~117B / ~5.1B active & Yes (MoE) & Aug 5, 2025 \\
\texttt{gpt-oss-20b} \cite{openai2025gptoss120bgptoss20bmodel} & ~21B / ~3.6B active & Yes (MoE) & Aug 5, 2025 \\
\texttt{gpt-oss-safeguard-120b} \cite{openai2025gptoss120bgptoss20bmodel} & ~117B / ~5.1B active & Yes (MoE) & Oct 2025 \\
\texttt{gpt-oss-safeguard-20b} \cite{openai2025gptoss120bgptoss20bmodel} & ~21B / ~3.6B active & Yes (MoE) & Oct 2025 \\
\texttt{Qwen3-235B-A22B-Instruct-2507} \cite{qwen3} & ~235B / ~22B active & Yes (MoE) & July 2025 \\
\bottomrule
\end{tabular}
\caption{Model parameter counts, MoE status, and release dates for selected LLMs.}
\label{tab:model_params_moe}
\end{table}

\subsection{Integration with the Detection Pipeline}

The resulting risk scores form the foundational lookup table for the pipeline’s risk aggregation step (Section~\ref{proposed-solution}).  
Each permission name is matched case-insensitively against the LLM-generated cache.  
When a permission is missing from the database, the system defaults to structural heuristics (floors/caps) based on regex pattern matching to maintain continuity of risk computation.

\section{Framework Overiew}
\label{sec:consent-detection-framework}

\begin{figure}[htb!]
  \centering
  \includegraphics[width=\linewidth]{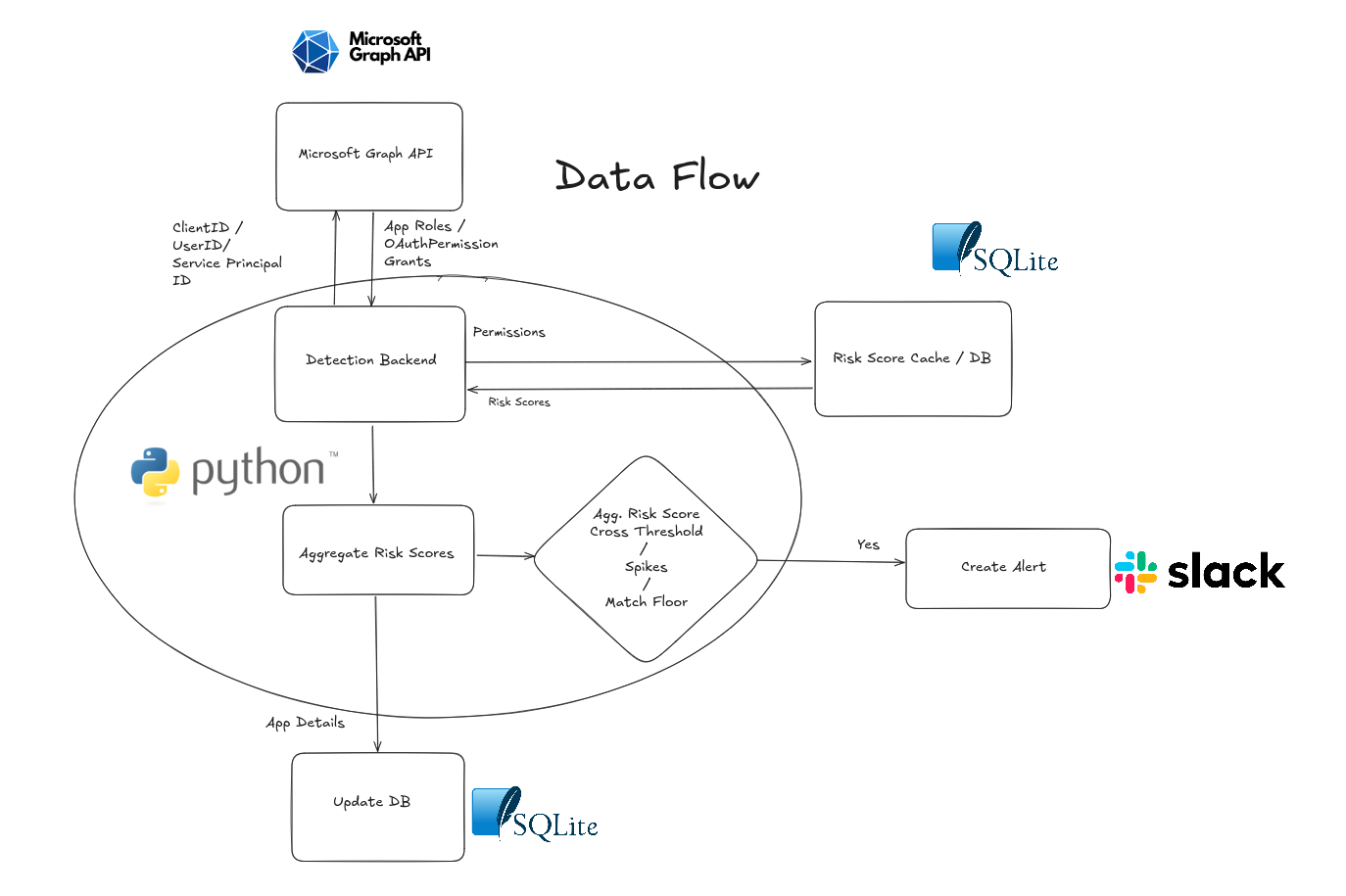}
  \caption{Rogue Application Detection Framework}
  \label{fig:consent-data-flow}
\end{figure}

Figure~\ref{fig:consent-data-flow} summarizes the detection framework we built to inventory OAuth consents and app role grants in a tenant, score their security risk, and generate actionable alerts. The design is intentionally opinionated around three operational requirements: (1) reliable collection from Microsoft Graph (internal vs external apps), (2) deterministic and explainable risk scoring of permissions, and (3) durable storage and alerting suitable for continuous monitoring.

\subsection{Data sources and collection (Microsoft Graph)}
The pipeline begins by querying Microsoft Graph for consent-related artifacts associated with an identity boundary such as a \emph{Client ID}, \emph{User ID}, or \emph{Service Principal ID}. \footnote{A full call tree is provided in Appendix \ref{appendix: graph api calls} for all the API calls needed in the data collection phase.} Two primary categories of authorization data are collected:

\begin{itemize}
  \item \textbf{OAuth delegated consent grants} (OAuthPermissionGrants): captures delegated permissions granted to an application on behalf of a user (or for all users, depending on grant type).
  \item \textbf{Application role assignments} (App Roles / AppRoleAssignments): captures app-only permissions or role assignments that can imply high-impact tenant access without interactive user context.
\end{itemize}

The Detection Backend normalizes raw Graph responses into a consistent internal representation (e.g., app identity, publisher/tenant context, consent type, principal scope, permission scope strings, and timestamps). This normalization step is critical because downstream scoring expects permissions to be canonicalized (space- vs comma-separated scopes, duplicates removed, stable ordering, etc.). 

The Detection Backend requires privileged access to the following \textbf{4} Graph API permissions to robustly collect the data required:

\begin{itemize}
    \item \textit{User.Read.All} allows the backend to collect information about all the users in a tenant.
    \item \textit{DelegatedPermissionGrant.Read.All} allows the backend to collect delegated grants for enumerated users.
    \item \textit{Directory.Read.All} allows the backend to collect the application catalogs for the enumerated users.
    \item \textit{Application.Read.All} allows the backend to collect the metadata for applications in the catalog such as \textit{Publisher}, and requested \textit{Permissions}.
    \item \textit{User.Read} is required for any application that needs to access Graph API.
\end{itemize}

While requesting for access via consent prompt, the backend is also assigned \textit{openid} and \textit{offline\_access} scopes since they are also essential for applications that require programmatic access to the Graph API.

\subsection{Risk scoring with caching (SQLite)}
Each discovered permission scope is assigned a \emph{base risk score} using a model- or policy-driven scoring function. To avoid rescoring the same permission repeatedly (and to enable reproducibility), the framework persists results in a \textbf{risk score cache} implemented as SQLite (Figure~\ref{fig:consent-data-flow}, ``Risk Score Cache / DB'').

Caching serves three practical goals:
\begin{enumerate}
  \item \textbf{Performance:} repeated tenant scans converge to mostly cache hits.
  \item \textbf{Stability:} scores for the same permission remain consistent across runs unless the scoring policy changes.
  \item \textbf{Auditability:} cached records can retain the score, the rationale, and the model/policy version used to compute it.
  \item \textbf{Spike Logic:} historical reference of permissions for the applications are saved for our spike detection; where we alert any spikes in requested/granted permissions for any application. \footnote{The \textit{last\_alerts} database table (DB Schema: Table \ref{tab:last-alerts-schema}) is used to reduce redundant alerts.}
\end{enumerate}

\subsection{Aggregate scoring and detection logic}
Permissions rarely appear in isolation; real-world applications request bundles of scopes, and risk emerges from both \emph{individual} high-impact scopes and \emph{combinations} that imply privilege escalation paths. The ``Aggregate Risk Scores'' stage computes an application-level (or grant-level) assessment from the set of permission scores. While the score aggregation follows our proposed solution from Section \ref{proposed-solution}, we explain our core detection logic, along with our detection pipeline, in detail in the next section (Section \ref{sec: detection pipeline}).

\subsection{Alerting (Slack) and persistence (SQLite)}
When the detection predicate evaluates to \textsf{Yes}, the framework emits an alert to Slack (Figure~\ref{fig:consent-data-flow}, ``Create Alert''). Alerts are designed to be triage-friendly: they include the app identity, the relevant application type (internal vs external), the specific permissions contributing to risk, the computed aggregate score, and the reasoning traces from the scoring function.

In parallel, the pipeline writes durable state to SQLite (Figure~\ref{fig:consent-data-flow}, ``Update DB'') including:
\begin{itemize}
  \item observed consents/grants and their normalized permission sets,
  \item application details (publisher, tenant, identifiers, metadata used for enrichment),
  \item the latest aggregate score and historical snapshots used for spike detection.
\end{itemize}
This persistence layer makes the system idempotent (re-runs do not duplicate alerts unnecessarily) and supports longitudinal analysis (e.g., ``what changed since last scan?''). \footnote{DB Schema available in Appendix: Table \ref{tab:applications-schema}}

\subsection{Summary}
In summary, the framework implements a full closed-loop detection system for OAuth consent risk: it collects grants from Microsoft Graph, scores and aggregates permissions with a cached and auditable SQLite layer, detects risky conditions using thresholds/spikes/floors, and notifies responders via Slack while persisting state for continuous monitoring (Figure~\ref{fig:consent-data-flow}).

\section{Detection Pipeline} \label{sec: detection pipeline}

The detection pipeline operationalizes the theoretical framework introduced in Section~\ref{proposed-solution}, embedding it into an automated telemetry--driven system that continuously inventories Microsoft Entra ID consent state and detects new or evolving permission grants. Figure~\ref{fig:pipeline-image} illustrates the dataflow between consent enumeration, permission resolution, risk aggregation, and alert generation.

\usetikzlibrary{arrows.meta, positioning, fit, backgrounds}

\tikzset{
  >={Latex[length=10mm]},
  stage/.style={draw, thick, fill=black!5, rounded corners=4pt, inner sep=4pt, minimum width=3.7cm, text width=3.9cm, align=left},
  title/.style={font=\bfseries\small\scshape, align=center},
  arrow/.style={-Latex, thick},
  note/.style={font=\scriptsize\itshape},
}

\begin{figure}[htbp]
\centering
\resizebox{\textwidth}{!}{%
\begin{tikzpicture}[node distance=1.0cm and 1.0cm, font=\footnotesize]

\node[title] (t1) at (0,5) {Data Sources};

\node[stage, below=2mm of t1] (audit) {Microsoft Graph (Directory Objects)\\
• \path|/applications| (\texttt{signInAudience=MyOrg})\\
• \path|/servicePrincipals| (all)\\
• \path|/users| (\texttt{\$select=id,UPN})};

\node[stage, below=5mm of audit] (apps) {Consent + Role Grant Artifacts\\
• \path|/users/{id}/oauth2PermissionGrants|\\
• \path|/oauth2PermissionGrants|\\
\hspace*{0.8em}\texttt{?\$filter=clientId eq \{spId\}}\\
• \path|/servicePrincipals/{id}/appRoleAssignments|};

\node[title, right=4.0cm of t1] (t2) {Ingestion};

\node[stage, below=2mm of t2] (collector) {ConsentCollector()\\
• \path|collect_internal_consents()|\\
\hspace*{0.8em}-- \path|requiredResourceAccess| (declared)\\
\hspace*{0.8em}-- delegated grants (\path|oauth2PermissionGrants|)\\
\hspace*{0.8em}-- app-only grants (\path|appRoleAssignments|)\\
• \path|collect_external_consents()|\\
\hspace*{0.8em}-- enumerate all \path|servicePrincipals|\\
\hspace*{0.8em}-- user+admin delegated + app roles};

\node[stage, below=5mm of collector] (resolver) {ExternalAppResolver()\\
• SP id $\rightarrow$ (appId, name)\\
• tenant-owned filter\\
\hspace*{0.8em}\path|/applications?| \texttt{\$filter=appId eq ...}\\
• caches: spId/appId};

\draw[arrow] (audit.east) -- ++(0.4,0) |- (collector.west);
\draw[arrow] (apps.east)  -- ++(0.4,0) |- (collector.west);

\node[title, right=4.0cm of t2] (t3) {Consent Resolution};
\node[stage, below=2mm of t3] (bag) {Permission Bag $\mathcal{P}$\\
• scopes (delegated)\\
• app roles (app-only)\\
• resolved names/values};
\node[stage, below=5mm of bag] (floors) {Floors \& Caps\\
• floor(5): *.ReadWrite.All, RoleMgmt.*, App/SP RW\\
• floor(4): *.Read.All, *.Send\\
• cap(2): AppFolder, createdByApp};

\draw[arrow] (collector.east) -- ++(0.4,0) |- (bag.west);
\draw[arrow] (resolver.east)  -- ++(0.4,0) |- (bag.west);

\node[title, right=4.0cm of t3] (t4) {Risk Scoring (GM, Tiers, Spikes)};
\node[stage, below=2mm of t4] (agg) {Risk Aggregation $R_{\text{app}}$\\
$s(p)=\max(r(p),f(p))$\\
$b=\max\{\mathrm{median},\max,M_p,\max f\}$\\
GM order $p=3$ (configurable)\\
Synergy: \texttt{offline\_access}+any $\ge3$ $\Rightarrow$ +1 tier};
\node[stage, below=4mm of agg] (tier) {Tier Mapping\\
critical $\ge$4.5, high $\ge$3.5, medium $\ge$2.0\\
cap tempering if no high floor};
\node[stage, below=4mm of tier] (spike) {Spike Logic (stateful)\\
$\theta=5$, count\_spike, ratio, cooldown=24h\\
multi/ratio spike $\Rightarrow$ alert};

\draw[arrow] (bag.east)    -- ++(0.4,0) |- (agg.west);
\draw[arrow] (floors.east) -- ++(0.4,0) |- (agg.west);
\draw[arrow] (agg.south)   -- (tier.north);
\draw[arrow] (tier.south)  -- (spike.north);

\node[title, right=4.0cm of t4] (t5) {Persistence \& Alerts};
\node[stage, below=2mm of t5] (alerts) {Threshold Alerts\\
• new / tier\_increase / perm\_added\\
• respects alert\_once\_per\_app};
\node[stage, below=4mm of alerts] (db) {State DB (SQLite)\\
• applications, run\_metadata, last\_alerts};
\node[stage, below=4mm of db] (slack) {SlackNotifier\\
• sends blocks + modifiers\\
(floors / caps / synergy)};

\draw[arrow] (tier.east)  -- ++(0.4,0) |- (alerts.west);
\draw[arrow] (tier.east)  -- ++(0.4,0) |- (db.west);
\draw[arrow] (spike.east) -- ++(0.4,0) |- (alerts.west);
\draw[arrow] (alerts.south) -- (db.north);
\draw[arrow] (db.south) -- (slack.north);

\coordinate (legendanchor) at (apps.south west |- resolver.south west);

\node[stage, fill=black!3, text width=1.0\linewidth, align=left,
      below=1.2cm of legendanchor, anchor=west] (legend) {
\textbf{Notes:} GM order default $p{=}3$; synergy bump triggers when \texttt{offline\_access} coexists with any permission of score $\ge3$;
cap tempering only applies when no high floors are present; spike alerts may bypass tier threshold depending on configuration.
};

\end{tikzpicture}%
}
\caption{Detection Pipeline Overview}
\label{fig:pipeline-image}
\end{figure}
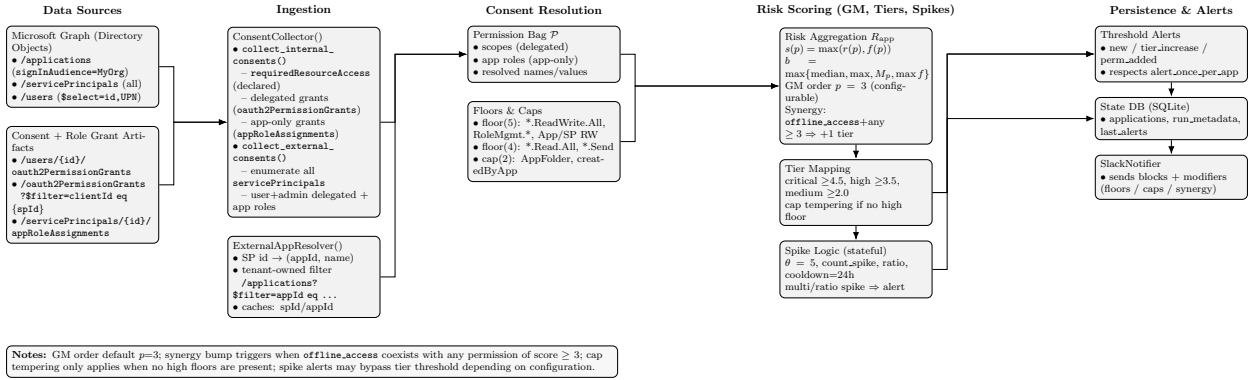

\subsection{Stage 1 — Consent and Application Enumeration}

The pipeline begins with a \texttt{ConsentCollector} module that interacts with Microsoft Graph to enumerate both internal (tenant-owned) and external (non-tenant-owned) applications and their associated grant artifacts.

\begin{itemize}
    \item \textbf{Internal Applications:} Queried via \texttt{/applications} (filtered to \texttt{signInAudience=AzureADMyOrg}) and augmented with:
    \begin{enumerate}
        \item \texttt{requiredResourceAccess} (declared permissions requested by the app, resolved to scope/role values via the resource service principal schema),
        \item \texttt{oauth2PermissionGrants} (delegated scopes granted to the app \emph{via} its service principal),
        \item \texttt{appRoleAssignments} (application permissions and other app roles granted to the app's service principal).
    \end{enumerate}

    \item \textbf{External Applications:} Enumerated \emph{first} from \texttt{/servicePrincipals} to ensure coverage of OIDC-only and ID-token-only apps and apps with no consent grants. The collector then enriches these entries by pulling:
    \begin{enumerate}
        \item \textbf{User-granted delegated scopes} from per-user \texttt{/users/\{id\}/oauth2PermissionGrants},
        \item \textbf{Admin-granted delegated scopes} from tenant-wide \texttt{/oauth2PermissionGrants?\$filter=clientId eq \{spId\}},
        \item \textbf{Application permissions (app roles)} from \texttt{/servicePrincipals/\{spId\}/appRoleAssignments}.
    \end{enumerate}

    The \texttt{ExternalAppResolver} dereferences service principal IDs to global \texttt{appId} and display name and filters out tenant-owned apps by checking whether an \texttt{/applications?\$filter=appId eq ...} result exists.
    This stage produces compact summaries \((\texttt{appId}, \texttt{displayName}, \mathcal{P}, U)\), where $\mathcal{P}$ is the union of granted scopes and app roles and $U$ is the set of observed consenting users (if any).
\end{itemize}

\subsection{Stage 2 — Risk Aggregation Runtime}

For every app instance, the detector queries a precomputed LLM-assisted permission risk cache (\texttt{permission\_analysis.db}) and applies the aggregation logic from Section~\ref{proposed-solution}:

\begin{itemize}
    \item Each permission has a base risk score and associated \textit{reasoning trace} as generated by an LLM.
    \item Each permission’s base risk is adjusted by applicable floors or caps.
    \item The resulting scores are combined using a generalized mean ($p\!=\!3$), together with median and max components, to yield $R_{\text{app}}$.
    \item The system records not just the aggregate score but also the modifiers applied (e.g., ``floor=5: RoleManagement.ReadWrite.Directory; synergy: offline\_access+$\ge3$'').
\end{itemize}

This enriched metadata enables both numerical and qualitative analysis during postmortem review.

\subsection{Stage 3 — Stateful Spike Logic and Temporal Memory}

Each run maintains persistent state in \texttt{detection\_state.db}, capturing the last known permission set, risk tier, and spike signature for each application.
Spike detection compares the new and previous high-risk permission sets:
\[
\Delta^+ = S_t - S_{t-1}, \quad \Delta^- = S_{t-1} - S_t,
\]
where $S_t = \{p \mid s(p)\ge 5\}$ at time $t$.
Alerts are generated when new spikes appear or when multiple p5-level permissions emerge simultaneously, subject to a configurable cooldown window (default: 24~hours).
This provides temporal stability and prevents alert storms.

\subsection{Stage 4 — Alert Generation and Reporting}

Alerts are dispatched via a structured \texttt{SlackNotifier} component.
Each alert block includes:
\begin{itemize}
    \item App identity and publisher domain,
    \item Risk tier and total score,
    \item Top-$k$ risky permissions with LLM rationales,
    \item Delta context (e.g., new permissions, tier increases, spike deltas),
    \item Risk modifiers applied (floors, caps, synergy).
\end{itemize}

Alert types follow a consistent taxonomy:
\begin{itemize}
    \item \texttt{new} — new application observed in the tenant.
    \item \texttt{tier\_increase} — risk tier escalation from prior snapshot.
    \item \texttt{perm\_added} — newly granted permissions detected.
    \item \texttt{spike\_present} / \texttt{spike\_multiple} — emergence of critical (p5) scopes.
\end{itemize}

\subsection{Stage 5 — State Update and Run Metadata}

At the end of each detection cycle, the pipeline:
\begin{itemize}
    \item Upserts each app’s current risk profile and permissions into the state database.
    \item Logs operational metadata such as total apps processed and number of new or changed apps.
\end{itemize}
This stateful persistence ensures that future runs can identify deltas over time, enabling long-term trend analysis and retrospective validation of detection efficacy.

\subsection{Summary}

In summary, this pipeline constitutes an end-to-end operationalization of the theoretical model from Section~\ref{proposed-solution}.
The LLM-driven risk cache supplies semantic weighting for each Graph permission, while the detector fuses directory-derived consent state (service principals, delegated grants, and app role assignments) with optional event-driven telemetry and temporal differencing to surface high-fidelity indicators of malicious app persistence within the tenant environment.

\section{Risk Score Analysis}

The foundation of the detection logic present in Section \ref{sec: detection pipeline} are the risk scores generated from the LLMs (listed in Table \ref{tab:model_params_moe}) using the methodology outlined in Section \ref{sec: llm-perm-risk-gen}. In this section we analyze the risk scores assigned by the LLMs to the permissions and share our analysis.\footnote{We have excluded the Llama and Granite models since they were unable to generate risk scores for all of the permissions (only labeled 740 and 457 permissions respectively) due to various reasons such as improper JSON formatting.}

\subsection{Everyone V\/S Everyone}
\subsubsection{Variations in Risk Scores}

There are some substantial variations in risk scores assigned by the 6 LLMs. Low risk permissions like \textit{profile} are labeled as medium risk (3) while high risk permissions such as \textit{TrustFrameworkKeySet.Read.All} (which allows tenant‑wide read access to trust framework key sets that can include private signing keys) are labeled as low risk (1).

\begin{table}[h!]
\centering
\begin{tabular}{lcccccc}
\hline
\textbf{permission\_name} & \textbf{mean} & \textbf{std} & \textbf{var} & \textbf{min} & \textbf{max} & \textbf{count} \\
\hline
Notifications.ReadWrite.CreatedByApp & 2.33 & 1.03 & 1.07 & 1 & 4 & 6 \\
APIConnectors.Read.All & 3.17 & 0.98 & 0.97 & 2 & 4 & 6 \\
offline\_access & 2.17 & 0.98 & 0.97 & 1 & 4 & 6 \\
TrustFrameworkKeySet.Read.All & 3.17 & 0.98 & 0.97 & 2 & 4 & 6 \\
TeamsAppInstallation.ReadWriteSelfForTeam & 2.83 & 0.98 & 0.97 & 1 & 4 & 6 \\
CallDelegation.Read & 2.00 & 0.89 & 0.80 & 1 & 3 & 6 \\
SensitivityLabel.Evaluate.All & 3.00 & 0.89 & 0.80 & 2 & 4 & 6 \\
IdentityProvider.Read.All & 3.00 & 0.89 & 0.80 & 2 & 4 & 6 \\
profile & 2.00 & 0.89 & 0.80 & 1 & 3 & 6 \\
Chat.Read.All & 3.50 & 0.84 & 0.70 & 3 & 5 & 6 \\
\hline
\end{tabular}
\caption{Permission risk statistics across models (rounded to two decimal places).}
\label{tab:permission_stats}
\end{table}

\begin{figure}[htb!]
    \centering
    \includegraphics[width=0.75\linewidth]{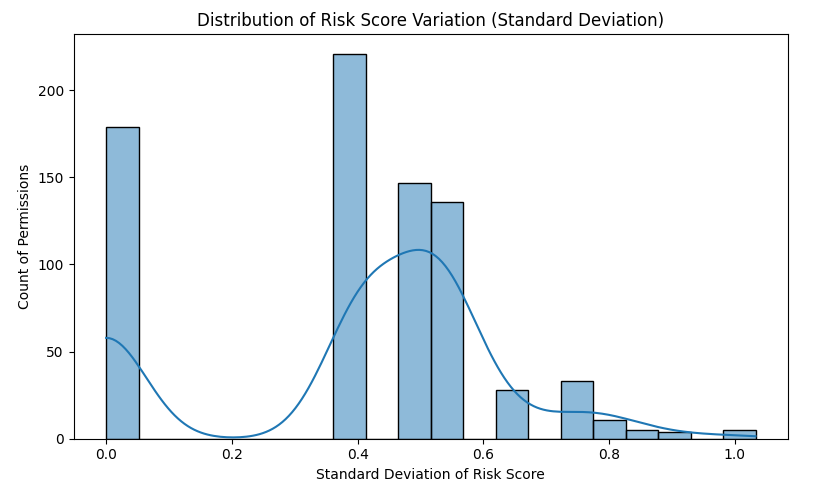}
    \caption{Standard Deviation of Scores}
    \label{fig:placeholder}
\end{figure}

\subsubsection{Distribution of Mean Scores}

Most of the permissions (\textbf{446}) are labeled as Medium to High Risk (3.0 - 4.0) while non-trivial (\textbf{50}) amount of permissions fall under low risk ($\leq$ 2.0).

\begin{figure}[htb!]
    \centering
    \includegraphics[width=0.75\linewidth]{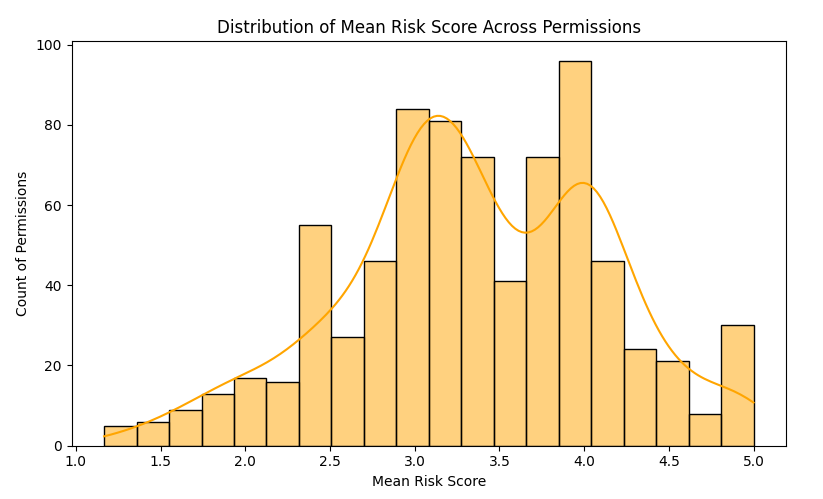}
    \caption{Distribution of Average Scores}
    \label{fig:dist_avg_scores}
\end{figure}

\subsection{GPT Family}

We also conducted a comparative analysis of risk scores between the different GPT models and their families; namely \textit{gpt-oss} and \textit{gpt-oss-safeguard} \cite{openai2025gptoss120bgptoss20bmodel}.

The GPT models largely agree on the risk scores (358 with no disagreements) for the permissions. Only 23 permissions have a difference of 2 (in risk score assignment) while 388 have a difference of 1; which is to say 388 permissions have 1 tier of difference while only 23 have 2 tier difference in risk assignment. 

\begin{figure}[htb!]
    \centering
    \includegraphics[width=0.75\linewidth]{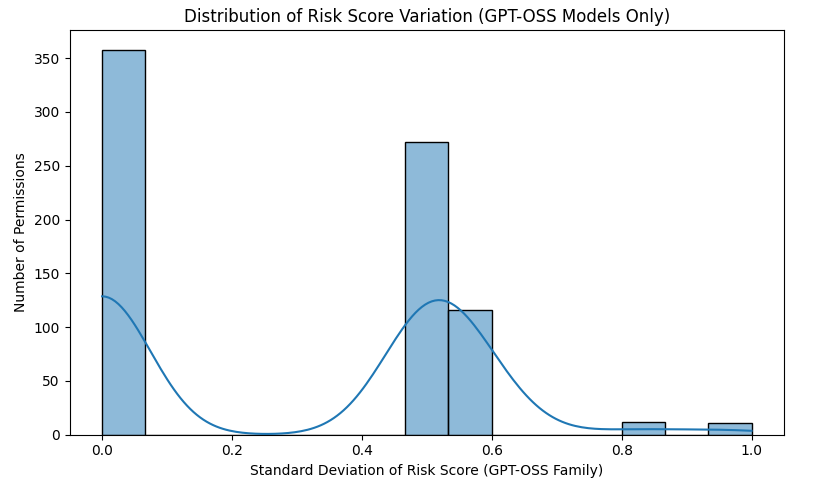}
    \caption{GPT OSS Variations}
    \label{fig:gpt_oss_variatons}
\end{figure}

\begin{figure}[htb!]
    \centering
    \includegraphics[width=0.75\linewidth]{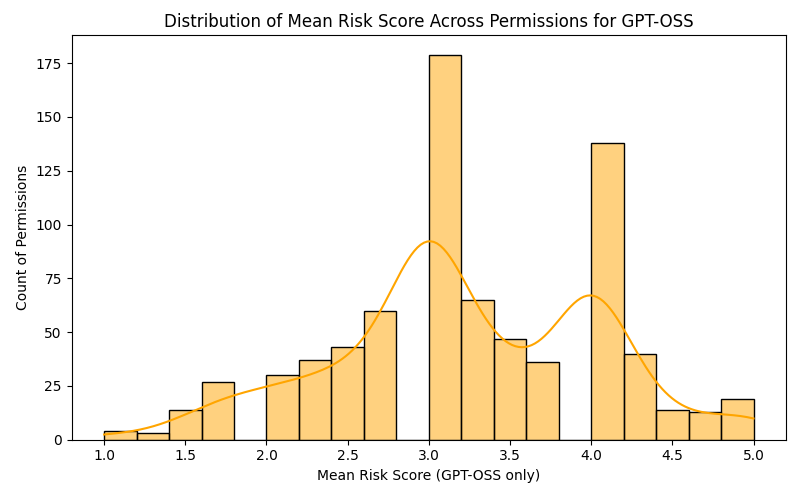}
    \caption{GPT-OSS Score Distribution}
    \label{fig:gpt_oss_mean_scores}
\end{figure}

The average scores from the \textit{gpt-oss} family clearly separates different permissions into distinctive tiers (as seen in Fig \ref{fig:gpt_oss_mean_scores}). A majority of the scores (327) were classified as medium risk while 224 permissions were classified as high-critical risk. 


\begin{figure}[htb!]
    \centering
    \includegraphics[width=0.75\linewidth]{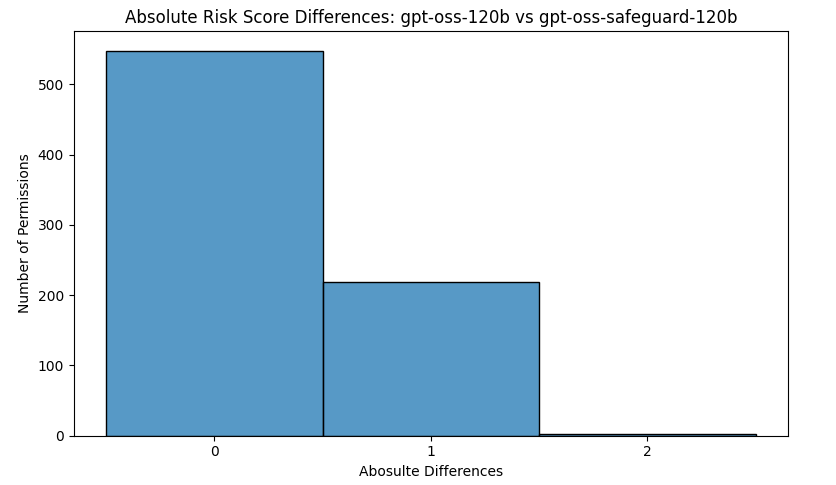}
    \caption{Absolute Differences between gpt-oss-120b and gpt-oss-safeguard-120b}
    \label{fig:abs-diff-gpt-oss}
\end{figure}

The \textit{gpt-oss-120b} and \textit{gpt-oss-safeguard-120b} seem to mostly agree on their risk assessments (Fig \ref{fig:abs-diff-gpt-oss}). Only 2 permissions were above 2 tiers of disagreements while 226 permissions were disagreed upon by a single risk tier.

\begin{figure}[htb!]
    \centering
    \includegraphics[width=0.75\linewidth]{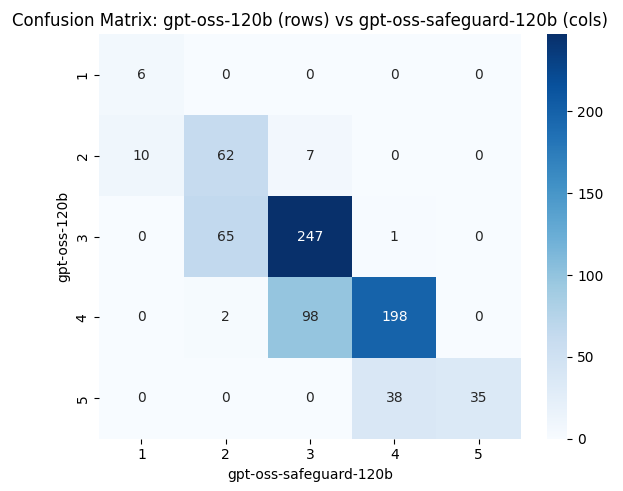}
    \caption{Confusion Matrix gpt-oss-120b vs gpt-oss-safeguard-120b}
    \label{fig:gpt-oss-confusion-matrix}
\end{figure}

We also observed that \textit{gpt-oss-safeguard-120b} consistently lowered the risk tier of permissions when compared to \textit{gpt-oss-120b} (Fig \ref{fig:gpt-oss-confusion-matrix}).

\subsection{Battle of Big Parameters}

We were also interested in comparing the results for the larger models only. We compared the outputs from the three big models: \textbf{Qwen3-235B-A22B-Instruct-2507}, \textbf{gpt-oss-120b} and \textbf{gpt-oss-safeguard-120b}.

We found that the \textit{Qwen} model generally aligns more with the vanilla \textit{gpt-oss} than \textit{gpt-oss-safeguard}. The \textit{gpt-oss-safeguard} consistenly lowers the risk tiers when compared to both vanilla \textit{gpt-oss} and the \textit{qwen} model.

\begin{figure}[htb!]
    \centering
    \includegraphics[width=1\linewidth]{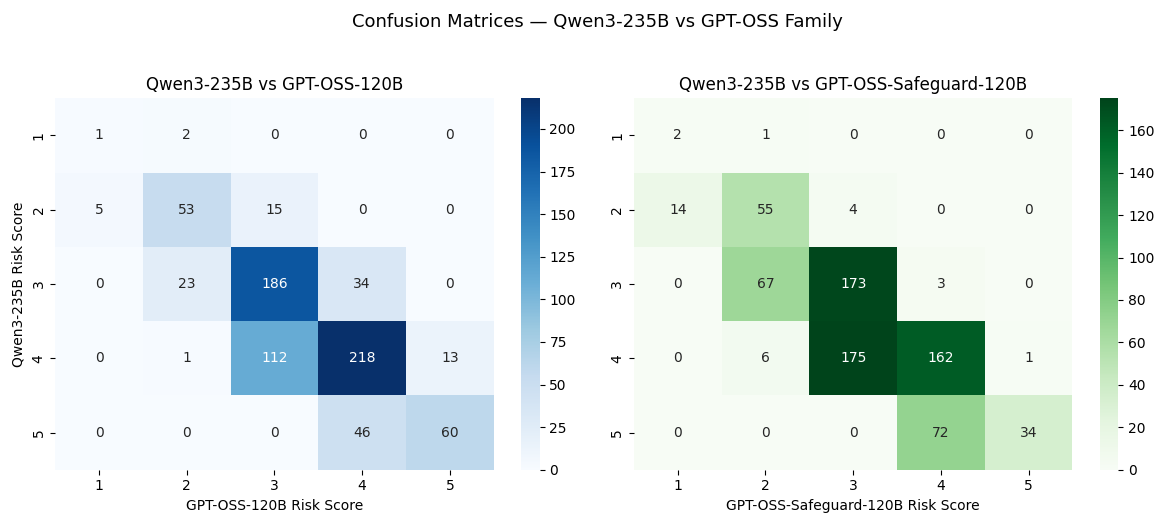}
    \caption{QWEN vs GPT}
    \label{fig:qwen_gpt_confusion}
\end{figure}

\subsubsection{Average Risk Scores}

\begin{figure}[htb]
    \centering
    \includegraphics[width=0.75\linewidth]{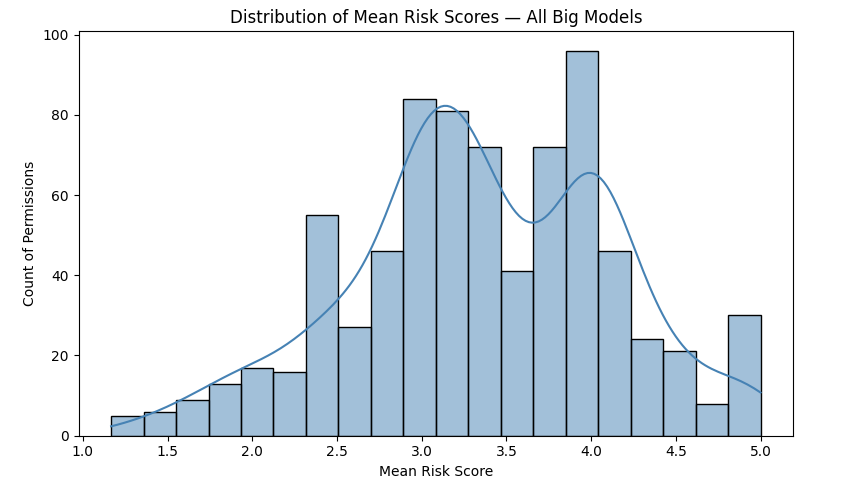}
    \caption{Average Scores (All Big Models)}
    \label{fig:mean_scores_all_big_models}
\end{figure}

Averaging the scores from all 3 big models gives us a dense distribution of the risk scores across the tiers. There is a larger concentration of risk scores in the medium risk region (3.0 - 4.0) than elsewhere with this approach (Fig \ref{fig:mean_scores_all_big_models}).

\begin{figure}[htb!]
    \centering
    \includegraphics[width=0.75\linewidth]{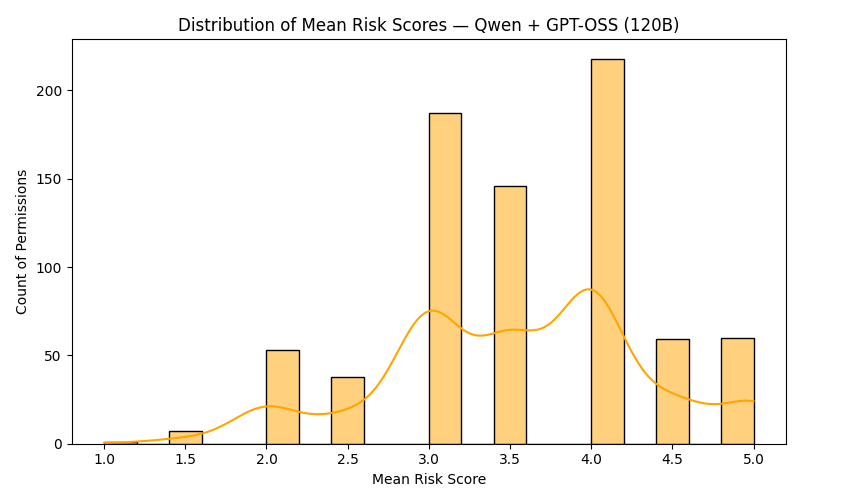}
    \caption{Average Scores (Qwen3 and GPT-OSS-120B)}
    \label{fig:mean_scores_qwen_oss_only}
\end{figure}

\begin{figure}[htb!]
    \centering
    \includegraphics[width=0.75\linewidth]{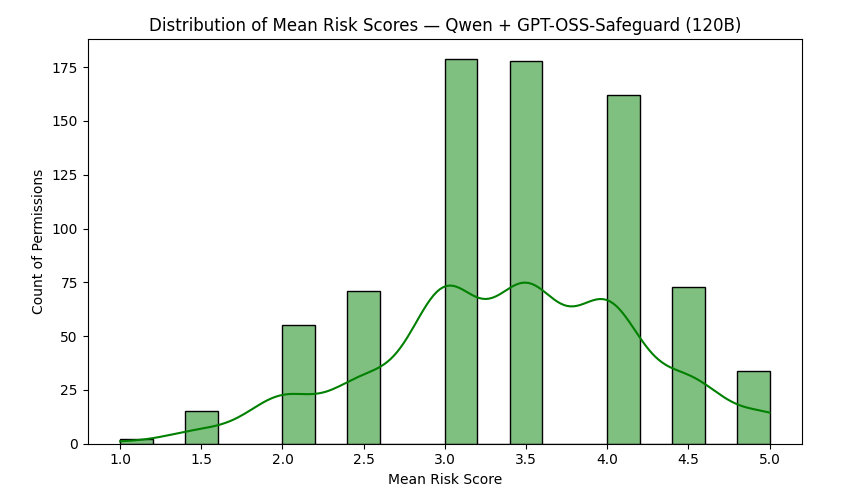}
    \caption{Average Scores (Qwen3 and GPT-OSS-SAFEGUARD-120B)}
    \label{fig:mean_scores_qwen_safeguard}
\end{figure}

Combining the \textit{qwen3} scores with the \textit{gpt-oss} results in conservatively higher risk tiers for a majority of the scores ( Fig \ref{fig:mean_scores_qwen_oss_only}) while combining with \textit{gpt-oss-safeguard} results in a medium-risk dense distribution of scores (Fig \ref{fig:mean_scores_qwen_safeguard}).

\section{N-grams Analysis of LLM Reasoning}

We also performed a short n-grams analysis for the reasoning generated by the LLMs for their respective risk score assignments. To ensure fair comparisons we remove any \textit{stop-words} (commonly occurring words like "the") from the reasoning texts by using the NLTK toolkit. \footnote{https://www.nltk.org/}

Instinctively, the models in the same family, such as \textit{gpt-oss}, and models derived from the same base model family such as \textit{Qwen3 and R1-Distill-Qwen}, have higher similarities than when compared to other models. We find that \textit{gpt-oss-120b} and \textit{gpt-oss-safeguard-120b} have the highest similarity for bi-grams while their smaller counterparts, \textit{gpt-oss-20b} and \textit{gpt-oss-safeguard-20b} have the highest similarity for tri-grams.

\begin{figure}[htb!]
    \centering
    \includegraphics[width=1.0\linewidth]{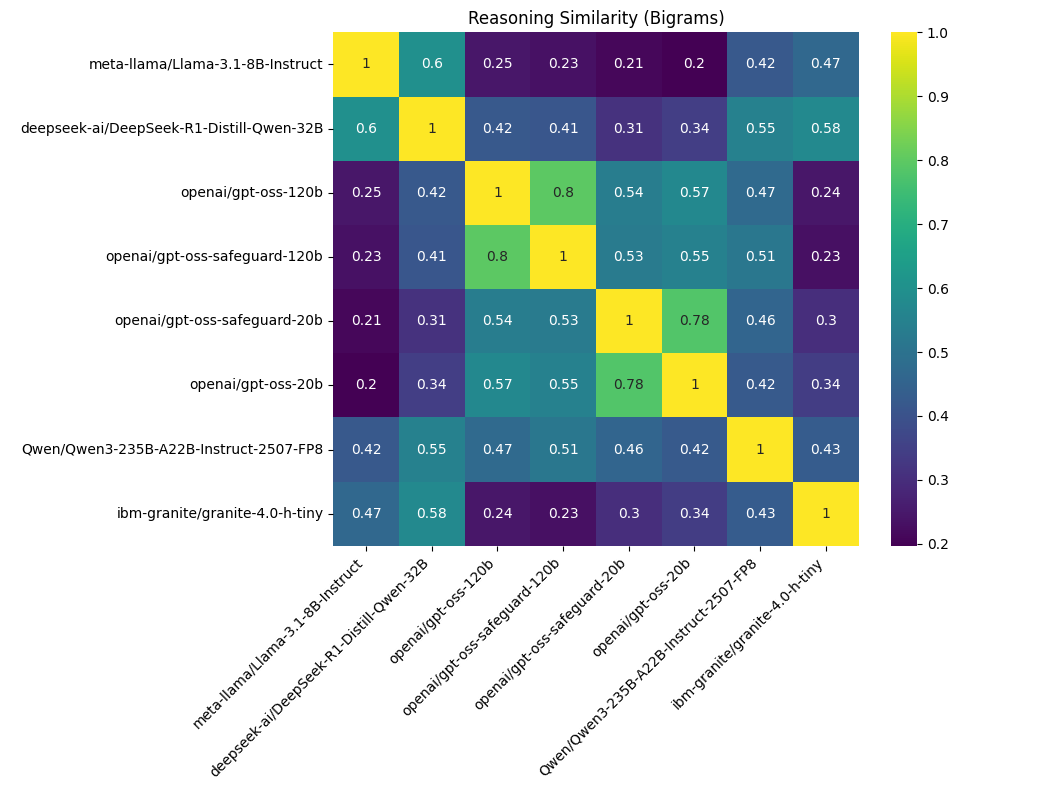}
    \caption{Bi-grams similarity for reasoning of permission risk scores from different LLMs}
    \label{fig:bigrams}
\end{figure}

\begin{figure}[htb!]
    \centering
    \includegraphics[width=1.0\linewidth]{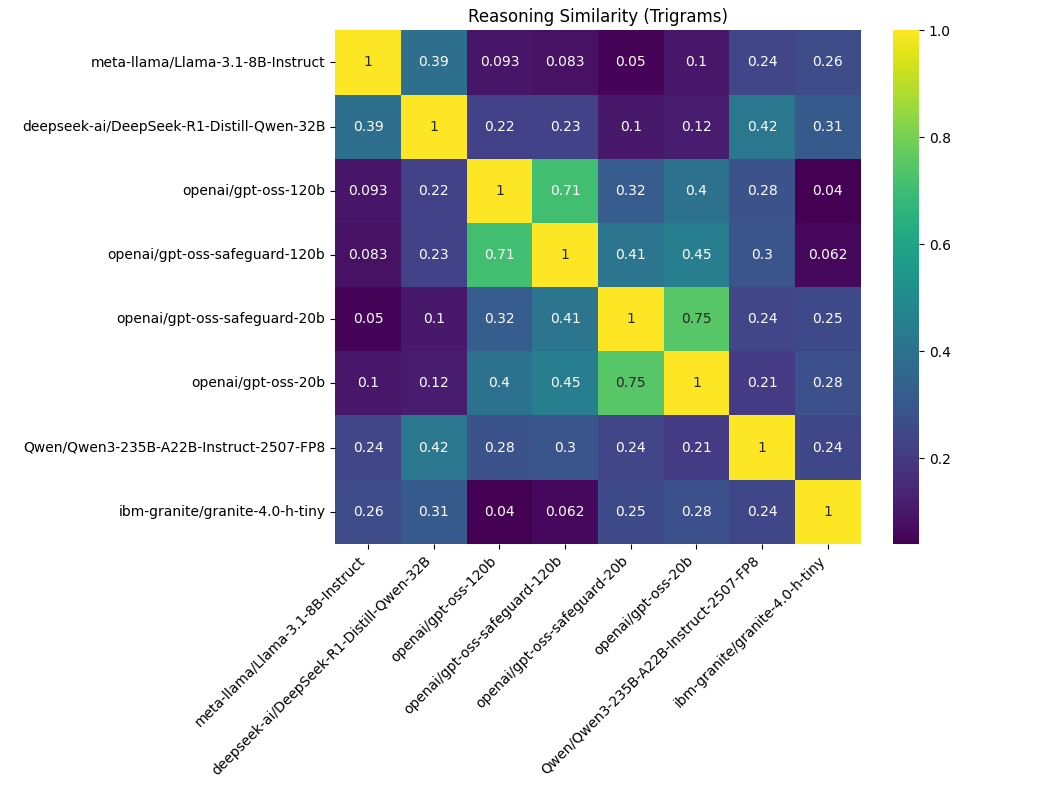}
    \caption{Tri-grams similarity for reasoning of permission risk scores from different LLMs}
    \label{fig:placeholder}
\end{figure}
\section{Prompt Engineering} \label{sec:prompt-engineering}

Prompt Engineering has been defined as the "means by which LLMs are programmed via prompts" \cite{prompt-engineering}. As prompt engineering has been increasingly recognized as an important part of LLM-based outputs [\cite{aws-prompt-engineering}, \cite{google-prompt-engineering}, \cite{ibm-prompt-engineering}, \cite{nextra-prompt-engineering}], we also performed a small experiment while changing the initial prompt for the risk scoring of the permissions.

\subsection{New Prompt Generation}

We prompted OpenAI GPT-5 with Deep Research \cite{openai-deep-research} feature enabled to improve our previous prompt. The exact prompt given for the improvement was:

\begin{boxA}\label{prompt-v0}
\small
You are a cybersecurity risk assessment specialist. Analyze the provided Microsoft Graph API permission and assign a comprehensive risk score from 1-5, where:

\begin{itemize}
    \item \textbf{1:} Minimal risk (public data, read-only)
    \item \textbf{2:} Low risk (limited read access)
    \item \textbf{3:} Moderate risk (broad read or limited write access)
    \item \textbf{4:} High risk (sensitive data write access)
    \item \textbf{5:} Critical risk (full administrative control)
\end{itemize}

\subsection*{Assessment Framework}
Evaluate the permission based on these risk factors:

\begin{enumerate}
    \item \textbf{Data Sensitivity \& Scope}
    \begin{itemize}
        \item What types of data can be accessed (PII, credentials, organizational data)?
        \item Is access limited to user's own data or broader organizational access?
    \end{itemize}
    
    \item \textbf{Privilege Escalation Potential}
    \begin{itemize}
        \item Could this permission be used to gain additional privileges?
        \item Does it enable modifying security settings or user permissions?
    \end{itemize}
    
    \item \textbf{Impact of Misuse}
    \begin{itemize}
        \item What damage could occur if maliciously used?
        \item Potential for data breach, service disruption, or financial impact?
    \end{itemize}
    
    \item \textbf{Attack Vectors}
    \begin{itemize}
        \item How might attackers exploit this permission?
        \item Common misuse patterns in real-world attacks?
    \end{itemize}
\end{enumerate}

\subsection*{Response Format}
Provide a JSON object with the following structure:

\begin{verbatim}
{
    "risk_score": 5,
    "reasoning": "Concise explanation covering key risk factors"
}
\end{verbatim}

\textbf{Permission Data:}
\\

make the above prompt better
\end{boxA}

Subsequent instructions to tune the prompt were:
\begin{itemize}
    \item make the variable json input in the last end so that prefix caching kicks in
    \item can you make the prompt even better? do some hardcore research and use all the tools and everything available to you.
\end{itemize}    
    
The resultant prompt, after deep research, was \textbf{4142} tokens long when compared to the previous prompt - which was \textbf{278} tokens long. The model cited 15 different references from resources such as \textit{learn.microsoft.com}, \textit{research.splunk.com}, \textit{tenable.com}, and \textit{uwconnect.uw.edu}. When reviewing the cited sources, all of them contained valuable information about different permissions, admin consents and official recommendations from Microsoft. The generated prompt can be found in Appendix \ref{appendix: full-llm-risk-prompt}.

\subsection{Impact of Prompt Engineering}

We applied the new prompt to generated risk scores from \textit{gpt-oss-120b} and \textit{Qwen3-235B-A22B-Instruct-2507} only.

\begin{figure}[htb]
    \centering
    \includegraphics[width=0.75\linewidth]{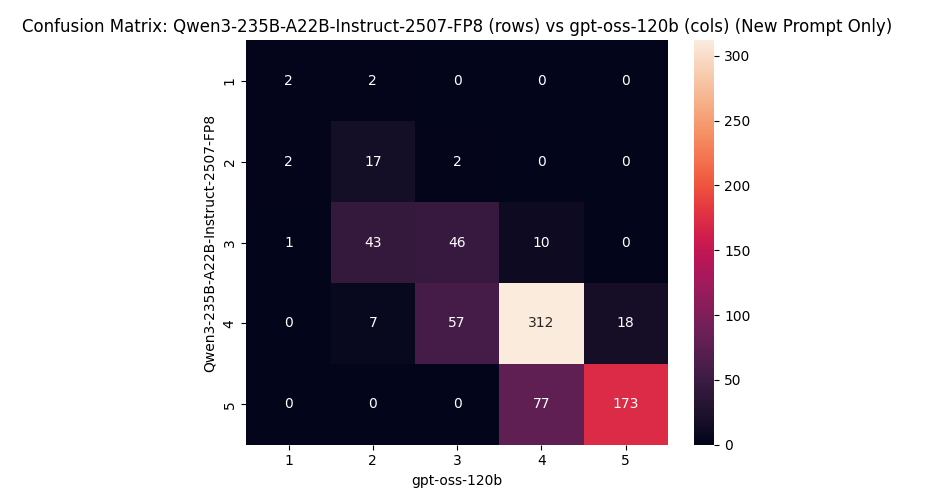}
    \caption{Qwen and GPT-OSS (with new prompt)}
    \label{fig:new-prompt-qwen-gpt}
\end{figure}

\begin{figure}[htb!]
    \centering
    \includegraphics[width=0.5\linewidth]{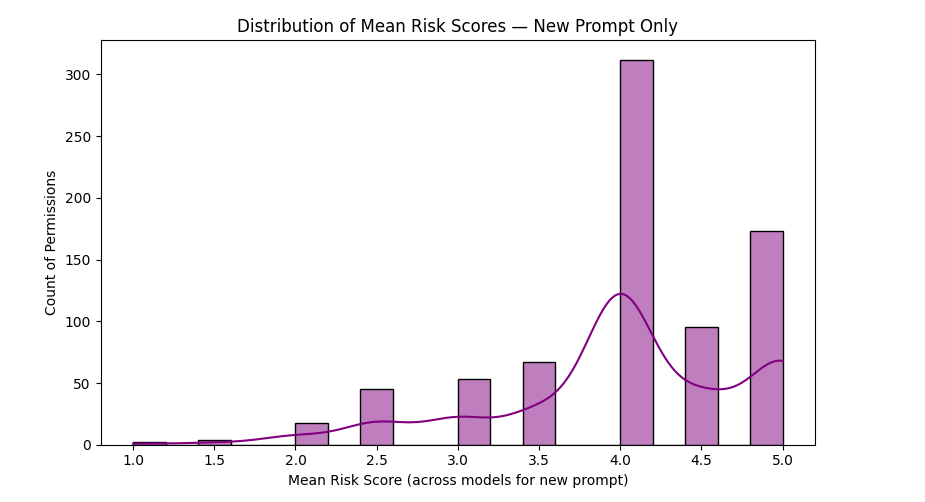}
    \caption{Average Scores (with new prompt)}
    \label{fig:mean_scores_new_prompt}
\end{figure}

We can see in Fig \ref{fig:new-prompt-qwen-gpt} that there are now less disagreements between the 2 models as when compared to the old prompt (as seen in Fig \ref{fig:qwen_gpt_confusion}). A significant increase in high-critical risk permissions can also be observed (Fig \ref{fig:mean_scores_new_prompt}). 

\subsubsection{Changes in GPT-OSS-120B Scores}

\begin{figure}[htb!]\label{fig:transition-matrix-v0-v1}
    \centering
    \includegraphics[width=0.5\linewidth]{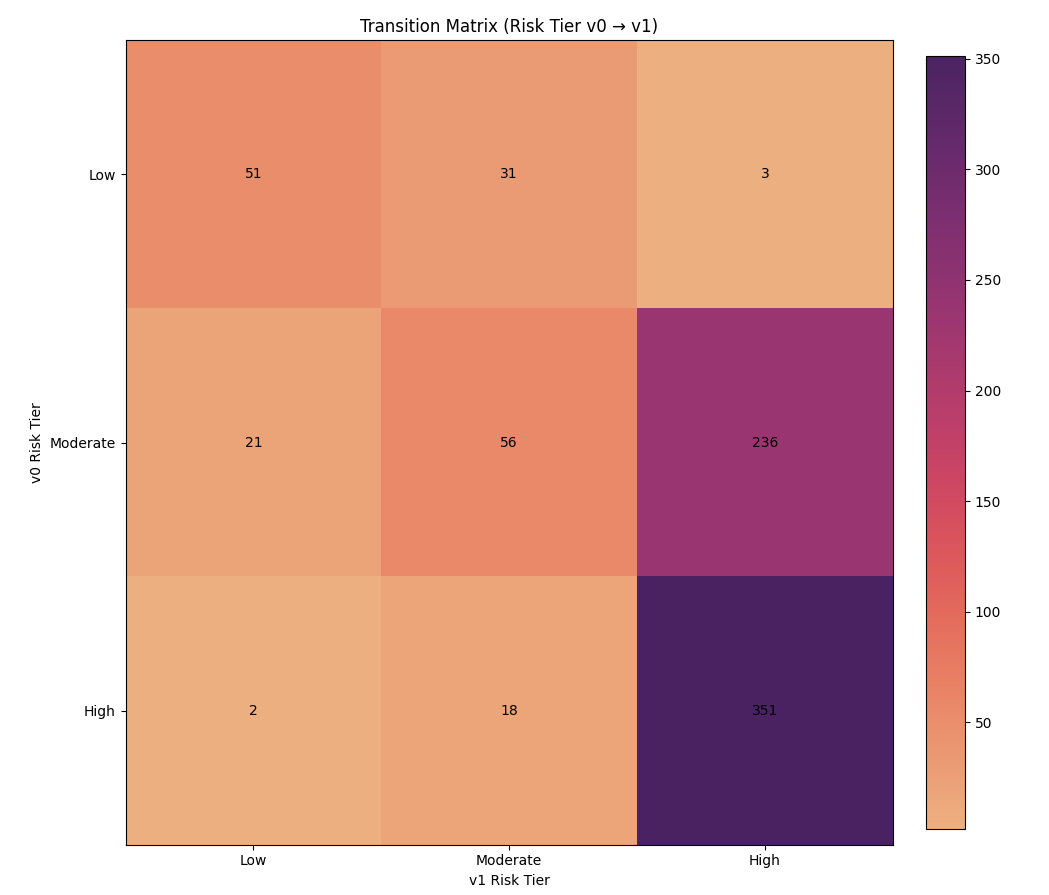}
    \caption{Changes in risk tiers from \textit{gpt-oss-120b} for new prompt}
    \label{fig:placeholder}
\end{figure}

We find that the scores from \textit{gpt-oss-120b} changed drastically with the new prompt. \textbf{394} permissions were scored into a higher tier while \textbf{47} permissions were scored lower and \textbf{328} permissions were scored in the same tier out of the total 769 permissions (as can seen in the transition matrix Fig \ref{fig:transition-matrix-v0-v1}). We contribute these changes to the instructions in the new prompt that instructed the model to assign higher risk scores in-case of confusion (more risk aversion) and included examples of low-risk permissions and scenarios in the new prompt (as seen in Appendix: \ref{appendix: full-llm-risk-prompt}). 

\begin{figure}[htb!]\label{fig:risk-tier-changes-bar-plot}
    \centering
    \includegraphics[width=0.5\linewidth]{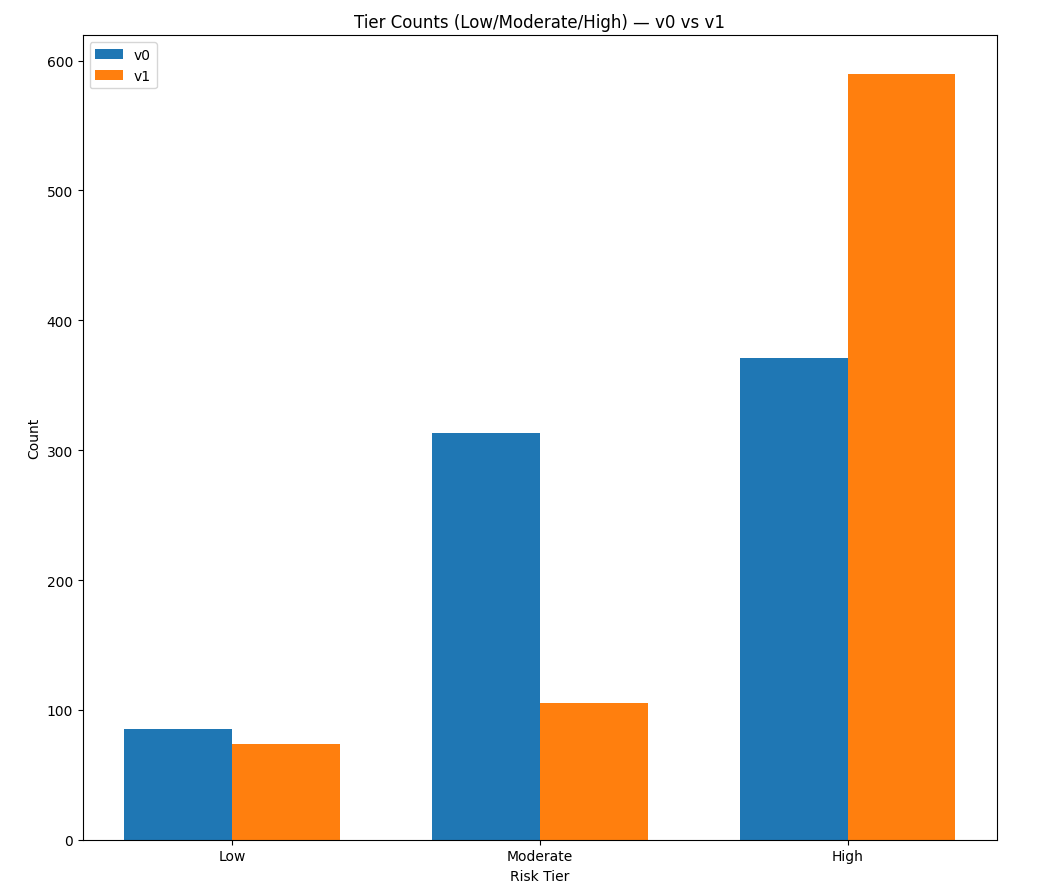}
    \caption{Changes in risk tier from \textit{gpt-oss-120b} for new prompt}
    \label{fig:placeholder}
\end{figure}

\section{Findings}

For this section, we strictly used the risk scores obtained by using the new prompt - as outlined in Section \ref{sec:prompt-engineering} - with the \textbf{gpt-oss-120b} model. \footnote{The model decision was arbitrary.}

\subsection{Tenant-Wide Permissions are Bad}

We found that permissions with \textbf{*.All} permissions were consistently scored higher by the models. The behavior can be intuitively understood since the \textbf{*.All} permissions give tenant-wide access to the application for the associated permission.

\begin{table}[h!]
\centering
\begin{tabular}{lcc}
\hline
\textbf{Risk Score} & \textbf{Count of Permissions} & \textbf{Percentage (\%)} \\
\hline
1 & 0 & 0.0 \\
2 & 10 & 2.0 \\
3 & 48 & 9.5 \\
4 & 298 & 59.0 \\
5 & 149 & 29.5 \\
\hline
\textbf{Total} & \textbf{505} & \textbf{100.0} \\
\hline
\end{tabular}
\caption{Distribution of Microsoft Graph permission risk scores}
\label{tab:risk_distribution}
\end{table}

Only the \textbf{*.All} permissions in the Table \ref{tab:readall_risk2} were labeled as \textbf{2} (low risk). The low risk read all permissions expose no PII; some are also classified by Microsoft as the lowest privilege alternative permission in their documentations \footnote{https://learn.microsoft.com/en-us/graph/api/industrydata-referencedefinition-list?view=graph-rest-beta}. 

\begin{table}[h!]
\centering
\begin{tabular}{ll}
\hline
\textbf{Permission Name} & \textbf{Risk Score} \\
\hline
Acronym.Read.All & 2 \\
Agreement.Read.All & 2 \\
IndustryData.ReadBasic.All & 2 \\
IndustryData-TimePeriod.Read.All & 2 \\
OrgSettings-Todo.Read.All & 2 \\
PrinterShare.ReadBasic.All & 2 \\
PrinterShare.Read.All & 2 \\
ProfilePhoto.Read.All & 2 \\
ServiceMessage.Read.All & 2 \\
ServiceHealth.Read.All & 2 \\
\hline
\end{tabular}
\caption{Sample permissions containing \texttt{.All} with risk score = 2}
\label{tab:readall_risk2}
\end{table}

\subsection{Share User with \textit{profile} Permission}

The safest and least-risky way to share basic user information is by using the \textbf{profile} permission; which is also classified as low-risk (2). The \textbf{profile} permission only shares name, picture and email of the user.

\subsection{Read-Only Permissions are Critically Risky Too}

The permissions in Table \ref{tab:readall_risk5} were classified as critical risk. All of the \textbf{*Read.All} permissions listed in the table allow access to critical identity information such as Certificates, Passwords and Authentication Mechanisms. 

\begin{table}[h!]
\centering
\begin{tabular}{ll}
\hline
\textbf{Permission Name} & \textbf{Risk Score} \\
\hline
BitlockerKey.Read.All & 5 \\
DeviceLocalCredential.Read.All & 5 \\
CustomAuthenticationExtension.Read.All & 5 \\
DeviceManagementCloudCA.Read.All & 5 \\
TrustFrameworkKeySet.Read.All & 5 \\
\hline
\end{tabular}
\caption{High-risk permissions containing \texttt{.Read.All} with risk score = 5}
\label{tab:readall_risk5}
\end{table}

\newpage
\subsection{Only Few Permissions are Minimal Risk}
Out of the 769 permissions, only 5 were classified as the lowest risk (as seen in Table \ref{tab:lowrisk_1}). 3 out of the 5 allow the application to read and write to Files, Notes and Notifications created by its own code; the other 2 include the programmatic access permissions: \textbf{offline\_access} and \textbf{openid}. The \textit{openid} permission allows an application to use \textit{OAuth} and \textit{OpenID Connect} for authentication while the \textit{offline\_access} permission allows an application to request \textit{refresh tokens} and maintain persistence access \footnote{https://learn.microsoft.com/en-us/entra/identity-platform/v2-protocols}.

However, as outlined in Section \ref{proposed-solution} and Section \ref{sec: detection pipeline}, we use a synergy mechanism to bump the risk level to a higher tier if an application requests or acquires persistent access to riskier permissions through \textit{offline\_access}. So, while the permission itself is not harmful, the ability for an intruder to maintain persistent access through the permission is and should be treated as higher risk.

\begin{table}[h!]
\centering
\begin{tabular}{ll}
\hline
\textbf{Permission Name} & \textbf{Risk Score} \\
\hline
Files.ReadWrite.AppFolder & 1 \\
Notes.ReadWrite.CreatedByApp & 1 \\
openid & 1 \\
offline\_access & 1 \\
Notifications.ReadWrite.CreatedByApp & 1 \\
\hline
\end{tabular}
\caption{Low-risk permissions with risk score = 1}
\label{tab:lowrisk_1}
\end{table}

\subsection{CreatedByApp is Usually Safe}

The permissions with suffix \textit{CreatedByApp} allow the application with access to data, user and activity created by application as seen in Table \ref{tab:createdbyapp}. The only moderately risky permission is the permission \textbf{UserNotification.ReadWrite.CreatedByApp} which allows the application send, read, update and delete notifications for any user in the tenant using the application which can be abused for phishing or user impersonation.

\begin{table}[h!]
\centering
\begin{tabular}{ll}
\hline
\textbf{Permission Name} & \textbf{Risk Score} \\
\hline
Notes.ReadWrite.CreatedByApp & 1 \\
Notifications.ReadWrite.CreatedByApp & 1 \\
UserTimelineActivity.Write.CreatedByApp & 2 \\
UserActivity.ReadWrite.CreatedByApp & 2 \\
UserNotification.ReadWrite.CreatedByApp & 3 \\
\hline
\end{tabular}
\caption{CreatedByApp permissions with varying risk scores (1–3)}
\label{tab:createdbyapp}
\end{table}

\section{Conclusion}
As enterprises continue migrating core collaboration and productivity workflows to cloud suites, Microsoft Entra ID (formerly Azure Active Directory) has become a high-value security boundary where third-party and first-party applications routinely broker access to sensitive mail, files, chats, calendars, and directory objects. While this extensibility enables significant productivity gains, it also creates a persistent over-privilege problem: applications frequently request broad Microsoft Graph permissions whose requested (and granted) scopes far exceeds what is required for their purported function, thereby expanding the attack surface for abuse through malicious registrations, deceptive consent prompts, or post-compromise permission escalation.

This work investigated the research question: \textit{Is it possible to automatically identify over-privileged/rogue applications in Microsoft Entra?} We answered this by proposing and implementing an automated, tenant-oriented detection approach that does \emph{not} depend on a centralized corpus of application descriptions. Instead, our method treats Graph permissions themselves as the primary signal and combines (i) systematic, explainable risk scoring of Microsoft Graph delegated scopes and application roles with (ii) deterministic aggregation logic and (iii) continuous collection of real consent and role-grant artifacts from Microsoft Graph. The outcome is an operational framework that can inventory application authorizations within a tenant, quantify risk in a repeatable manner, and surface suspicious applications with analyst-ready reasoning.

Concretely, we produced a permission-centric risk dataset covering 769 Microsoft Graph permissions, where each permission is assigned an ordinal risk score and accompanied by a natural-language rationale. We then integrated these permission-level assessments into a detection pipeline that continuously collects delegated consent grants and app-only role assignments, enriches each authorization with app metadata, and computes an application-level score using conservative guardrails (floors/caps and tier mapping) as well as temporal and contextual amplifiers (e.g., persistence-aware synergy and spike detection). This design supports both transparency and operational stability: risk decisions are explainable, cacheable, and resilient to transient noise, while still sensitive to abrupt privilege changes that commonly occur in attack chains.

Overall, the results demonstrate that a permission-first approach can meaningfully approximate the notion of \emph{over-privilege} in Microsoft Entra without requiring complete application-intent data. By encoding Graph permission semantics into a reproducible risk policy and applying it directly to tenant authorization states, the system can highlight applications whose granted access is plausibly excessive or dangerous given the risk assessments of individual permissions. Importantly, the framework is engineered for continuous monitoring rather than one-time auditing, aligning detection with how OAuth abuse unfolds in practice: gradually accumulated access, long-lived refresh capability, and sudden escalations to tenant-wide privileges.

\subsection{Limitations}
This study has several limitations that motivate further work. First, while permission semantics are highly informative, over-privilege is ultimately contextual: the same permission set may be appropriate for one application and suspicious for another. Because our approach does not rely on comprehensive app descriptions, it prioritizes \emph{risk severity} over \emph{intent mismatch}, which may increase false positives for legitimate administrative or compliance tools. Second, Microsoft Graph visibility and tenant policy constraints (e.g., conditional access, limited audit retention, incomplete telemetry for some app behaviors) can affect collection completeness, which in turn impacts detection coverage. Third, LLM-assisted permission scoring introduces sensitivity to prompting and model choice; although we constrain outcomes with deterministic aggregation and store reasoning traces for auditability, score calibration may still drift over time as models and documentation evolve. Finally, the evaluation of ``true'' over-privilege remains hard without robust ground truth labels and incident-confirmed malicious applications, limiting evaluation to indirect validation and qualitative analysis.

\newpage

\subsection{Future Work}
Future work can extend this research in several directions:

\begin{itemize}
    \item \textbf{Ground truth and validation:} building a partially labeled benchmark (e.g., confirmed malicious consents, sanctioned enterprise apps, and known-good admin tools) would enable rigorous measurement of detection accuracy and false-positive rates.
    \item \textbf{Context enrichment:} incorporating lightweight intent signals---such as publisher reputation, verified domains, app category, sign-in patterns, and permissions \emph{usage} telemetry where available---could better distinguish necessary high privilege from suspicious excess.
    \item \textbf{Explainability:} integrating feature-attribution and counterfactual explanations could translate ``high risk'' into concrete remediation guidance (e.g., the minimal set of permissions driving the score, recommended least-privilege alternatives, and expected impact of revocation).
    \item \textbf{Policy and continuous updates:} versioning the scoring rubric and automatically tracking Graph permission lifecycle changes (new permissions, deprecations, behavioral changes) would improve long-term maintainability and support governance workflows.
    \item \textbf{Broader threat coverage:} extending the pipeline to correlate authorization risk with downstream indicators (service principal changes, credential additions, suspicious sign-ins, and mailbox/file access anomalies) could provide higher-confidence detections and earlier intervention opportunities.
    
\end{itemize}

In summary, this project demonstrates that automated identification of overprivileged applications in Microsoft Entra is feasible using a permission-centric lens. By converting Graph permissions into an explainable, operational risk policy and continuously applying it to tenant consent and role-grant state, defenders can obtain scalable visibility into one of the most abused and least-understood control planes in modern enterprise SaaS environments.

\printbibliography
\appendix
\section*{Appendix X: Full LLM Risk Assessment Prompt} \label{appendix: full-llm-risk-prompt}

\begin{tcolorbox}[
  title=Updated LLM Prompt Used for Microsoft Graph Permission Risk Scoring,
  colback=white,
  colframe=black,
  breakable,
  enhanced,
  sharp corners,
  boxrule=0.5pt
]

\begin{lstlisting}
You are an enterprise cybersecurity risk assessor specializing in Microsoft Graph API permissions. Your knowledge spans Microsoft's documentation and security best practices for Graph delegated and application permissions, consent models, and OAuth attack techniques. You evaluate a given Graph permission's risk impact on an organization, considering factors like data sensitivity, scope of access, privilege escalation potential, persistence/abuse avenues, and impact of misuse. Leverage Microsoft's current guidance and classifications for Graph permissions (e.g. high vs. low impact) and known attacker tactics to inform your assessment.
Always apply the principle of least privilege and assume a conservative security stance (if in doubt, err toward higher risk).

Risk Assessment Criteria

    Consider each factor below for the given permission, then determine an overall risk score. Permissions with broader scope, higher sensitivity, or privilege impact should be scored higher (more risky). Use Microsoft's definitions of high-impact permissions and known OAuth threats as guidance:

        Data Sensitivity: Identify the type of data or resource the permission grants access to. Permissions allowing access to highly sensitive data (e.g. user emails, mailbox contents, files/documents, audit logs, user credentials, or security policies) indicate higher risk. For example, reading mail or files can expose confidential information, so a permission like Mail.Read (user mail) is sensitive. Access to audit logs or directory data might reveal security configurations or personal data. More sensitive data = higher risk.
            - For example, reading mail or files can expose confidential information, so a permission like Mail.Read (user mail) is sensitive. Access to audit logs or directory data might reveal security configurations or personal data. More sensitive data = higher risk.

        Scope of Access: Determine if the permission is user-scoped or tenant-wide. Broad "*.All" scopes (or permissions that apply to all users or all items in the tenant) dramatically increase risk.
            - For instance, Files.Read.All (read all files in SharePoint/OneDrive across the org) or Mail.ReadWrite.All (access all mailboxes) affect the entire tenant's data and are extremely high risk.
            - In contrast, permissions confined to a single user's data (e.g. reading one user's mailbox or calendar) are less risky. Delegated vs. Application: Application permissions inherently operate tenant-wide without a user context, making them broadly powerful (usually requiring admin consent) and thus often riskier than equivalent delegated scopes.
            - A delegated permission applies only to the signed-in user's accessible data and actions; if that user has limited privileges, the impact is contained
            - (However, if a highly privileged user or admin token is used, even delegated permissions can reach many resources.) Always increase the risk score for any permission with global/All scope or tenant-wide impact, especially in application form.

        Privilege Escalation Risk: Check if the permission can be abused to gain higher privileges or persist in the environment. Some permissions allow managing directory roles, users, or applications-these are very dangerous because an attacker could leverage them to escalate privileges or create backdoors. Examples:

        Role or Directory Management: RoleManagement.ReadWrite.Directory lets an app assign roles or modify role assignments in Azure AD, potentially elevating itself or others to admin

            App Management: Application.ReadWrite.All or ServicePrincipal.ReadWrite.All allows creating or modifying application registrations and credentials. Attackers can inject credentials or add permissions to an app (or new malicious apps), achieving persistent privileged access.

            AppRole/Consent Grants: AppRoleAssignment.ReadWrite.All is extremely high risk - it can grant any permission to any app, bypassing normal consent flow and directly enabling admin-level access to resources.

            Policy or Settings Control: Permissions that let you alter security policies or tenant settings (e.g. conditional access policies, Intune device policies) can weaken defenses or create persistence, thus high risk. If a permission enables modifying sensitive configurations or assignments, consider it critical risk due to potential privilege escalation or persistence (even if it doesn't directly expose data).

        Persistence & Abuse Potential: Determine if the permission facilitates long-term access or covert abuse. Notably, the offline_access scope (often requested with delegated permissions) allows an app to maintain access continuously via refresh tokens, even when the user is offline.
            This means an attacker who gains a token with offline_access can persist access without further user interaction. While offline_access by itself doesn't grant new data access, it amplifies the risk of whatever other permissions are granted by enabling ongoing use (so it should raise the overall score for a permission set). Also consider if a permission could be abused in subtle ways - e.g. sending mail (Mail.Send) might not exfiltrate data, but an attacker could send fraudulent emails from a trusted account (potentially severe impact). Permissions that allow adding credentials or creating accounts (persistence mechanisms) should be rated higher. Any capability that could let an attacker remain in the environment or quietly expand their access contributes to a higher risk.

        Impact if Misused: Assess the worst-case scenario if an attacker maliciously exploits this permission. Ask: Could it lead to a major data breach? Facilitate lateral movement across many accounts or services? Allow the attacker to maintain a foothold even if user credentials are reset? The broader and more sensitive the potential impact, the higher the risk score. For example, a permission that allows reading all users' emails could expose a treasure trove of confidential data (high impact breach), while one that only reads a user's basic profile or their To-Do tasks has limited impact. Consider organizational damage: breach of sensitive data, impersonation of users, or disabling of security controls all warrant higher scores.

        Known Attacker Tactics: Be aware of common OAuth attack patterns. Consent phishing attacks often trick users (or admins) into granting malicious apps permissions like email or file access to harvest data. Attackers also target high-privilege app permissions to quietly exfiltrate data or add backdoors - for example, nation-state actors have used illicit app consents to gain mailbox access at scale and persist beyond the initial compromised account. If the permission in question is one frequently seen in attacks (for instance, Mail.Read, Mail.ReadWrite.All, Files.Read.All, or app-only Exchange full_access_as_app), treat it as higher risk due to its abuse in real-world breaches. Additionally, if a delegated permission normally only allows user-level access, but in the hands of an admin user it could grant tenant-wide data access (e.g. Mail.Read when used by a global admin could read many mailboxes), take that into account. Use Microsoft's security guidelines and documentation to justify your reasoning whenever possible.

Scoring Rubric (1-5)

    Using the factors above, map the permission to a risk level 1 through 5. Below are guidelines for each score with typical examples:

        Risk Score 5 (Critical) - The permission poses an extremely high risk to the enterprise. It likely grants broad, admin-level control or full access to highly sensitive data across the tenant. These permissions, if misused, could directly lead to catastrophic impact such as full tenant compromise, massive data breach, or creation of persistent backdoors.

            Characteristics: Tenant-wide "All" scope with write or full control, or the ability to escalate privileges or control security-critical settings. Often requires admin consent by design.

            Examples:
            - Application permissions like Directory.ReadWrite.All (read/write all directory objects) or User.ReadWrite.All (modify all user accounts) - can affect every user or group in the directory.
            - Mail.ReadWrite.All or Files.ReadWrite.All - read and modify all users' emails or files in the organization (huge data exposure plus ability to alter or destroy data).
            - AppRoleAssignment.ReadWrite.All - can assign any role to any app, effectively granting itself or another application high privileges (direct path to privilege escalation).
            - RoleManagement.ReadWrite.Directory - full control over directory roles (attacker could make themselves or an app a Global Admin).
            - Application.ReadWrite.All / ServicePrincipal.ReadWrite.All - create or update any enterprise application, including adding credentials or permissions (can establish persisting backdoors in apps).
            - Permissions to modify critical policies (e.g. conditional access, device compliance) or audit logs - could disable security or cover attack traces.

            Impact: If an attacker obtains a token for any of these, the entire tenant is at risk - they could exfiltrate or delete all organizational data, create rogue admin accounts, or permanently maintain access. Score 5 indicates maximum severity.

        Risk Score 4 (High) - The permission is highly sensitive and poses a major risk, though slightly more limited than a 5. It often grants broad read access to sensitive data or moderate write abilities. Misuse could cause significant damage or data loss but maybe without full administrative takeover.

            Characteristics: Could be a tenant-wide read-only permission for very sensitive data, or a permission with wide scope that lacks only a small aspect of the most critical permissions. Might also be a combination of factors (e.g. not "All" scope but still affecting many users or critical data). Usually requires admin consent.

            Examples:
            - Files.Read.All (application or delegated with admin consent) - read all files in all SharePoint sites and OneDrives. This exposes virtually all documents in the organization (sensitive intellectual property, financials, etc.).
            - Mail.Read.All - read all users' email messages. This can yield confidential communications and sensitive attachments, though read-only (still a major breach risk).
            - AuditLog.Read.All or SecurityEvents.Read.All - access to audit logs or security events for the whole tenant. While not personal data, these contain sensitive info about system configuration and user activities; an attacker could learn security measures or find weaknesses.
            - Delegated permission with "All" scope under an admin context, e.g. Delegated Mail.Read or MailboxSettings.ReadWrite used by an administrator account. By itself a delegated permission may only target the signed-in user's data, but if that user is an admin, it effectively grants org-wide impact (an admin's mailbox might contain high-level info, or an admin could access others' data).
            - User.Read.All (application) - read detailed profile of every user in the tenant. Exposes organizational directory data (potential privacy concern, list of all users, roles, etc.), though less sensitive than emails/files.
            - Group.Read.All - read info and membership of all Microsoft 365 Groups. Notably, this can include reading group content like Teams messages or files in those groups, which can be highly sensitive.
            - Mail.Send (if application permission to send as any user) or similar - ability to impersonate users in sending mail. This can be abused for spearphishing or fraud at scale, though it doesn't directly read data. (If only delegated Mail.Send for the signed-in user, impact is limited to that user's identity - likely lower risk.)

            Impact: A score of 4 means a major security threat: large-scale data exposure or significant account abuse is possible. It may not give full control over Azure AD or devices, but it substantially violates confidentiality or can facilitate targeted attacks (e.g. reading all mail for intel, or sending emails as executives). Organizations should tightly restrict these permissions and treat apps requesting them with extreme caution.

        Risk Score 3 (Moderate) - The permission carries a moderate risk. It typically grants access to sensitive data or actions, but in a limited scope (e.g. single user scope or a specific subset of data), or it grants broader access to less-sensitive data. Misuse could cause harm to an individual user's data or a segment of the org, but would not be an existential threat to the entire tenant by itself.

            Characteristics: Often delegated user-level permissions to important data, or application permissions to moderately sensitive data. The scope is usually one user at a time (for delegated) or a constrained set. User consent is sometimes allowed for these (if deemed low-impact by admin policy), but they still deserve caution.

            Examples:
            - Mail.Read or Mail.ReadWrite (Delegated, user-specific) - an app can read a single user's mailbox (with that user's consent). This is sensitive (user's email privacy) but impact is localized to that user's data. A compromised app could read or send email from one account (unless many users individually fall victim to consent phishing).
            - Files.Read (Delegated) - read files in the signed-in user's OneDrive or SharePoint sites they have access to. Could expose sensitive files for that user, but not others' files unless shared.
            - Calendars.ReadWrite (Delegated) - read/write the user's calendar. Could be used to surveil or alter one person's schedule or meetings (privacy/security concern but limited scope).
            - User.ReadBasic.All (Delegated) - read basic profile info of all users (name, email, etc.). This affects the whole directory but only exposes limited, non-confidential details (often considered a low-impact permission by Microsoft). Still, an attacker could enumerate the employee list for targeting, so it's more than minimal risk but not highly sensitive.
            - Contacts.Read (Delegated) - read a user's contacts. Leakage of personal contacts is a concern but not critical org data.
            - Group.Read.All (Delegated, user context) - if a regular user consents, the app can only see groups that user is a member of or can access (not all groups unless the user is an admin). So actual access is narrower than the name suggests.
            - Device.Read.All (Application) - read basic info of all devices in Azure AD. Reveals inventory data (which could help an attacker, but not as sensitive as user data).

            Impact: Score 3 indicates a noticeable risk: a breach or misuse could compromise one user's sensitive data or some subset of information. It could lead to targeted attacks (e.g. reading one executive's email could be very damaging in that context), but it wouldn't automatically compromise the entire organization. Still, such permissions should be granted sparingly and monitored, especially if multiple moderate permissions are combined or if the affected user has elevated privileges.

        Risk Score 2 (Low) - The permission has limited scope or low sensitivity, posing a relatively low risk to the organization. These permissions either provide access to non-critical data or actions, or are heavily constrained in what they can do. Even if abused, the potential harm is minor or confined.

            Characteristics: Likely read-only access to non-sensitive data, or write access to very benign settings. Possibly requires no admin consent (allowed for user consent by default) due to being considered low impact.

            Examples:
            - User.Read (Delegated) - read the signed-in user's basic profile (name, email, tenant ID). This is standard and exposes minimal info (often required for any app sign-in).
            - profile, email, openid - standard OpenID Connect scopes to get user identity info (name, email address) and authentication. These reveal only basic identity data and are needed for sign-in.
            - MailboxSettings.Read (Delegated) - read a user's mailbox settings (like email reply settings). Not emails themselves, just configuration; low impact in isolation.
            - Tasks.Read - read a user's To-Do or Planner tasks. Potentially sensitive regarding that user's work, but generally low impact if limited to one user.
            - User.Read.All (Delegated) - read full profile of all users that the signed-in user can access. For a normal user, that might just be themselves and maybe their contacts; it doesn't actually retrieve every user unless the signer is privileged. So impact is usually low. (If an admin grants it and uses it, it could read all user profiles, but user profile info is still low sensitivity compared to emails or files.)
            - Any permission restricted to a single user's less-sensitive data. For instance, Notes.Read (if it existed) for one user's notes.

            Impact: Score 2 indicates a minor risk. An attacker with this access could maybe gather some info about the organization or a user, but not secret or critical data. The damage would likely be limited and containable. Organizations can allow these permissions more freely (often pre-approved for user consent), though they should still be tracked.

        Risk Score 1 (Minimal) - The permission is effectively minimal risk. It either exposes virtually no sensitive information or is a very common baseline permission with no significant security impact. Even if misused, it would not meaningfully harm the user or organization.

            Characteristics: Often these are default or utility permissions needed for basic app functionality, or they grant access to information that is public or non-sensitive. They might also be static permissions with no user-specific data (e.g. access to a service's status).

            Examples:
            - openid, profile, email (when considered individually) - these reveal the user's identity basics and are required for authentication flows (considered safe and Low impact by Microsoft).
            - Offline_access by itself - this only allows the app to receive refresh tokens to keep session alive
            . While it enables persistence, if no high-impact permissions accompany it, on its own it doesn't grant new data access. (It should be scored higher in combination with other scopes, but as a standalone scope it's part of normal sign-in.)
            - User.Read - read one's own profile (redundant with openid/profile, minimal extra info).
            - Calendars.Read (Delegated) - read a signed-in user's calendar free/busy info. Arguably could be sensitive (meetings details), but many orgs consider calendar info low sensitivity by default sharing settings.
            - Public or quasi-public info access: e.g. a permission to read service health status or retrieve one's organization's public facing data.
            - Any permission explicitly classified as "Low impact" by admin policy.

            Impact: A score of 1 means negligible risk. The permission is very unlikely to be exploited for any meaningful malicious gain. It might be necessary for basic app operations and carries no confidential data access. These are typically safe to grant broadly.

Note: These examples are guidelines - in practice, evaluate the permission's context. If a typically low-risk permission is requested alongside many others or by an untrusted app, the overall risk might increase. Conversely, if a high-risk permission is constrained by additional controls or least-privilege design, its effective risk could be somewhat mitigated (though still treated with caution). Always default to the most sensitive interpretation of a permission when scoring, to avoid underestimating risk.

Additional Guidance

    Conservative Scoring: If unsure, choose a higher risk score. Permissions that include any form of "All" or write/admin capabilities should default to a high score (4 or 5) due to their breadth. Only truly low-impact, limited permissions should get 1 or 2. This ensures we don't under-rate a potentially dangerous permission.

    Justify Briefly: In the output JSON's "reasoning", include 1-2 sentences explaining why that score was chosen, referencing key factors (e.g. "has tenant-wide email access - high breach impact" or "read-only single-user data - limited scope"). The reasoning should be concise but specific to the permission.

    Reference Microsoft Guidance: Where relevant, base your reasoning on Microsoft's documented classification or warnings. For example, if Microsoft documentation or security reports label a permission as high risk or requiring admin consent, reflect that. (E.g., "Permission requires admin consent and grants global access to sensitive data, which Microsoft flags as high impact.") This adds credibility and context to your assessment.

    Attacker's Perspective: Always think how an attacker could abuse the permission. If it can be abused in known attack patterns (consent grant attacks, token replay, privilege escalation, data exfiltration), incorporate that into the risk reasoning. E.g., mention if "attackers could leverage this to maintain persistent access (using refresh tokens)" or "to impersonate users and phish others," etc. This helps explain the risk in practical terms.

Finally, proceed to assess the given permission.

Provide a JSON object with the following structure:

    "risk_score": <integer 1-5>,
    "reasoning": "<short explanation of risks>"

The permission to evaluate will be provided as a JSON object below.
\end{lstlisting}
\end{tcolorbox}

\section*{Appendix XI: Graph API calls for collecting consents}\label{appendix: graph api calls}

\begin{figure}[ht]
\centering
\scalebox{0.78}{ 
\begin{tikzpicture}[
    node distance=7mm and 18mm,
    process/.style={
        rectangle, rounded corners,
        draw, thick, align=center,
        fill=black!5,
        inner sep=5pt,
        text width=4.3cm    
    },
    arrow/.style={-Latex, thick}
]

\node[process] (sps) {1. Fetch \\All Service Principals\\(/servicePrincipals)};
\node[process, below=of sps] (class) {2. Determine Internal\\vs External Apps};

\node[process, below left=22mm and 26mm of class] (int4) {3A. Internal App Objects\\(/applications)};
\node[process, below=of int4] (int5) {4A. Declared Permissions\\\texttt{requiredResourceAccess}};
\node[process, below=of int5] (int6) {5A. Delegated Grants\\(/oauth2PermissionGrants?\\clientId=\dots)};
\node[process, below=of int6] (int7) {6A. App Role Assignments\\(/servicePrincipals/\{spId\}/\\appRoleAssignments)};
\node[process, below=of int7] (int8) {7A. Build Internal\\App Inventory};

\draw[arrow] (class.west) |- (int4);
\draw[arrow] (int4) -- (int5);
\draw[arrow] (int5) -- (int6);
\draw[arrow] (int6) -- (int7);
\draw[arrow] (int7) -- (int8);

\node[process, below right=22mm and 26mm of class] (ext_users) {3B. Fetch Users\\(/users)};
\node[process, below=of ext_users] (ext4) {4B. External SPs Detected\\(multi-tenant / OIDC apps)};
\node[process, below=of ext4] (ext5) {5B. User Delegated Grants\\(/users/\{id\}/\\oauth2PermissionGrants)};
\node[process, below=of ext5] (ext6) {6B. Admin Delegated Grants\\(/oauth2PermissionGrants?\\clientId=\dots)};
\node[process, below=of ext6] (ext7) {7B. External App\\Role Assignments\\(/servicePrincipals/\{spId\}/\\appRoleAssignments)};
\node[process, below=of ext7] (ext8) {8B. Build External\\App Inventory};

\draw[arrow] (class.east) |- (ext_users);
\draw[arrow] (ext_users) -- (ext4);
\draw[arrow] (ext4) -- (ext5);
\draw[arrow] (ext5) -- (ext6);
\draw[arrow] (ext6) -- (ext7);
\draw[arrow] (ext7) -- (ext8);

\draw[arrow] (sps) -- (class);

\end{tikzpicture}
} 
\end{figure}

\section*{Appendix XII: Qualitative Comparison}\label{appendix: qualitative comparison}

\setlength\LTleft{0pt}
\setlength\LTright{0pt}
\setlength\LTcapwidth{\textwidth}

\begingroup
\footnotesize
\setlength{\tabcolsep}{4pt}
\renewcommand{\arraystretch}{1.15}

\begin{xltabular}{\textwidth}{%
  >{\centering\arraybackslash}p{0.06\textwidth}%
  >{\raggedright\arraybackslash}p{0.23\textwidth}%
  >{\raggedright\arraybackslash}X%
  >{\raggedright\arraybackslash}X%
}
\caption{Comparative analysis of gpt-oss-120b and Qwen3-235B on reasoning from new prompt}
\label{tab:qual-compare}\\

\toprule
Risk & Permission & Reasoning (gpt-oss-120b) & Reasoning (Qwen3-235B) \\
\midrule
\endfirsthead

\toprule
Risk & Permission & Reasoning (gpt-oss-120b) & Reasoning (Qwen3-235B) \\
\midrule
\endhead

\midrule
\multicolumn{4}{r}{\footnotesize Continued on next page}\\
\endfoot

\bottomrule
\endlastfoot

1 & \makecell[l]{\texttt{Notes.ReadWrite.}\\\texttt{CreatedByApp}} &
The permission is deprecated and explicitly documented as having no effect; it grants no actual access to notes or other data, so it poses minimal or no security risk. &
The permission 'Notes.ReadWrite.CreatedByApp' is explicitly deprecated and no longer grants any functional access, as stated in Microsoft's description. It has no impact on data or privileges, even if consented to, making it effectively minimal risk. \\

\midrule
2 & \texttt{Acronym.Read.All} &
Acronym.Read.All only provides read-only access to a low-sensitivity resource (acronyms). Even though the application permission is tenant-wide, it cannot modify data or elevate privileges, so the potential impact of misuse is limited to exposure of non-critical information. &
The permission grants read access to acronyms, which are likely non-sensitive organizational terms or abbreviations. It requires admin consent for application use but is limited in sensitivity, posing low risk even with tenant-wide scope. Microsoft does not classify this as a high-impact permission, and it does not expose personal, confidential, or security-critical data. \\

\midrule
3 & \texttt{Files.Read} &
Files.Read is a delegated permission that lets an app read the signed-in user's OneDrive and SharePoint files. While file contents can be highly sensitive, the scope is limited to the individual user's accessible files and does not provide tenant-wide access. It requires user consent only, so the impact is confined to a single account, placing it in the moderate (score 3) risk category. &
The permission grants access to read files in the signed-in user's OneDrive and SharePoint sites they can access, which may include sensitive personal or business data. While scoped to a single user (delegated), misuse could lead to significant data exposure for that user, especially if they have access to confidential documents, warranting a moderate risk score. \\

\midrule
4 & \texttt{User.Read.All} &
User.Read.All provides tenant-wide read access to every user's full profile (including personal details, manager relationships, and custom attributes). This broad, admin-consented scope enables large-scale reconnaissance and targeted social-engineering attacks, making it a high-impact permission despite being read-only. &
This permission grants tenant-wide access to read all users' full profiles, including sensitive directory data such as reports, managers, and extended properties, which can aid in reconnaissance and targeted attacks. It requires admin consent and is classified by Microsoft as a high-impact permission due to its broad scope and potential to expose organizational structure and user relationships, enabling lateral movement and social engineering—especially if abused by an application or privileged user. \\

\midrule
5 & \texttt{Channel.Delete.All} &
Channel.Delete.All grants tenant-wide delete rights for Teams channels (application) and can delete any channel the signed-in user can access (delegated). This is an admin-consent, write-level permission that can cause massive data loss, erase evidence, and disrupt collaboration across the entire organization, matching Microsoft’s high-impact classification. &
This permission grants tenant-wide ability to delete channels in any team, both via application and delegated contexts, requiring admin consent. It has a broad 'All' scope and enables destructive actions across Microsoft Teams, potentially disrupting collaboration, erasing data, and enabling denial-of-service at scale. Microsoft classifies such write/delete operations with global scope as high-impact due to their irreversible impact and potential for abuse in sabotage or persistence by attackers. \\

\end{xltabular}
\endgroup

\section*{Appendix XIII: Database Schemas}\label{appendix: db_schema}

\begin{table}[htb!]\label{table:db_schema_permission_analysis}
\centering
\small
\caption{Schema: \texttt{permission\_analysis}}
\label{tab:permission-analysis-schema}
\begin{tabularx}{\textwidth}{@{} l l c l X @{}}
\toprule
\textbf{Column} & \textbf{SQL type} & \textbf{Null?} & \textbf{Default} & \textbf{Description} \\
\midrule
id & INTEGER & NO & AUTOINCREMENT & Internal primary key. \\
permission\_name & TEXT & NO & & Canonical name of the permission / scope being analyzed. \\
risk\_score & INTEGER & NO & & Numeric risk score assigned by the model. \\
model\_name & TEXT & NO & & Identifier of the model/version used to evaluate the permission. \\
reasoning & TEXT & YES & & Optional textual explanation / rationale produced by the model. \\
created\_at & TIMESTAMP & YES & CURRENT\_TIMESTAMP & Timestamp when the analysis record was created. \\
\midrule
\multicolumn{5}{@{}l@{}}{\textbf{Constraints / Indexes}} \\
\multicolumn{5}{@{}l@{}}{UNIQUE(permission\_name, model\_name) — prevents duplicate analyses for the same permission and model.} \\
\bottomrule
\end{tabularx}
\end{table}

\begin{table}[htb!]
\centering
\small
\caption{Schema: \texttt{applications}}
\label{tab:applications-schema}
\begin{tabularx}{\textwidth}{@{} l l c l X @{}}
\toprule
\textbf{Column} & \textbf{SQL type} & \textbf{Null?} & \textbf{Default} & \textbf{Description} \\
\midrule
id & INTEGER & NO & AUTOINCREMENT & Internal primary key. \\
app\_id & TEXT & NO (UNIQUE) & & Application identifier (unique). \\
display\_name & TEXT & YES & & Human-friendly name. \\
publisher\_domain & TEXT & YES & & Publisher domain / tenant. \\
type & TEXT & YES & & Classification (``internal'' | ``external''). \\
total\_risk & REAL & YES & & Numeric aggregate risk score. \\
risk\_level & TEXT & YES & & Categorical risk level. \\
permissions & TEXT & YES & & Permissions stored as JSON array (text). \\
last\_seen & TIMESTAMP & YES & CURRENT\_TIMESTAMP & Last-observed timestamp. \\
\bottomrule
\end{tabularx}
\end{table}

\begin{table}[htb!]
\centering
\small
\caption{Schema: \texttt{run\_metadata}}
\label{tab:run-metadata-schema}
\begin{tabularx}{\textwidth}{@{} l l c l X @{}}
\toprule
\textbf{Column} & \textbf{SQL type} & \textbf{Null?} & \textbf{Default} & \textbf{Description} \\
\midrule
id & INTEGER & NO & AUTOINCREMENT & Internal primary key. \\
run\_time & TIMESTAMP & YES & CURRENT\_TIMESTAMP & Timestamp when the run was recorded. \\
total\_apps & INTEGER & YES & & Total number of apps observed in the run. \\
new\_apps & INTEGER & YES & & Count of newly discovered apps in the run. \\
changed\_apps & INTEGER & YES & & Count of apps that changed since last run. \\
\bottomrule
\end{tabularx}
\end{table}

\begin{table}[htb!]
\centering
\small
\caption{Schema: \texttt{last\_alerts} (spike alert cool-down state)}
\label{tab:last-alerts-schema}
\begin{tabularx}{\textwidth}{@{} l l c l X @{}}
\toprule
\textbf{Column} & \textbf{SQL type} & \textbf{Null?} & \textbf{Default} & \textbf{Description} \\
\midrule
app\_id & TEXT & NO & & Primary key linking to application identifier. \\
last\_spike\_ts & TEXT & YES & & Timestamp of the last detected spike (stored as text). \\
last\_spike\_sig & TEXT & YES & & Signature/details of the last spike event. \\
\bottomrule
\end{tabularx}
\end{table}

\begin{table}[htb!]
\centering
\small
\caption{DB Indices}
\label{tab:indices}
\begin{tabularx}{\textwidth}{@{} l X @{}}
\toprule
\textbf{Index} & \textbf{Purpose / Notes} \\
\midrule
\texttt{idx\_apps\_last\_seen} & Non-unique index on \texttt{applications(last\_seen)} for time-based queries and lookup performance. \\
\texttt{idx\_apps\_appid} & Unique index on \texttt{applications(app\_id)} (enforces uniqueness and speeds app lookups). \\
\bottomrule
\end{tabularx}
\end{table}

\end{document}